\documentclass[sigconf,authorversion,nonacm,screen]{acmart}
\AtBeginDocument{%
  \providecommand\BibTeX{{%
    \normalfont B\kern-0.5em{\scshape i\kern-0.25em b}\kern-0.8em\TeX}}}

\usepackage{soul}
\usepackage{tabularx}
\usepackage{makecell}
\usepackage{multirow}

\copyrightyear{2024}
\acmYear{2024}
\setcopyright{rightsretained}
\acmConference[CHI EA ’24]{Extended Abstracts of the CHI Conference on Human Factors in Computing Systems}{May 11--16, 2024}{Honolulu, HI, USA}
\acmBooktitle{Extended Abstracts of the CHI Conference on Human Factors in Computing Systems (CHI EA ’24), May 11--16, 2024, Honolulu, HI, USA}
\acmDOI{10.1145/3613905.3651001}
\acmISBN{979-8-4007-0331-7/24/05}

\begin{document}

\title[Kawaii Computing]{Kawaii Computing: Scoping Out the Japanese Notion of Cute in User Experiences with Interactive Systems}

\author{Yijia Wang}
\email{wang.y.cf@m.titech.ac.jp}
\orcid{0009-0004-2250-9163}
\affiliation{%
  \institution{Tokyo Institute of Technology}
  \city{Tokyo}
  \country{Japan}
}

\author{Katie Seaborn}
\email{seaborn.k.aa@m.titech.ac.jp}
\orcid{0000-0002-7812-9096}
\affiliation{%
  \institution{Tokyo Institute of Technology}
  \city{Tokyo}
  \country{Japan}
}

\renewcommand{\shortauthors}{Wang and Seaborn}

\begin{abstract}
  Kawaii computing is a new term for a steadily growing body of work on the Japanese notion of ``cute'' in human-computer interaction (HCI) research and practice. Kawaii is distinguished from general notions of cute by its experiential and culturally-sensitive nature. While it can be designed into the appearance and behaviour of interactive agents, interfaces, and systems, kawaii also refers to certain affective and cultural dimensions experienced by culturally Japanese users, i.e., kawaii user experiences (UX) and mental models of kawaii elicited by the socio-cultural context of Japan. In this scoping review, we map out the ways in which kawaii has been explored within HCI research and related fields as a factor of design and experience. We illuminate theoretical and methodological gaps and opportunities for future work on kawaii computing.
\end{abstract}





\maketitle

\section{Introduction}

"Kawaii" is the Japanese term for "cute," simply put. Kawaii is recognized the world over as representative of Japanese popular culture~\cite{nittono2010behavioral}. Not limited in Japan, kawaii has been adopted as a global phenomenon across various forms of media and beyond~\cite{1}. It plays a critical role in the worldwide success of many Japanese products~\cite{160} and cultural exports~\cite{kinsella2013cuties,lieber2019word}, such as Hello Kitty~\cite{kovarovic2011hello} and the Pokémon series of anime and games~\cite{allison2003portable}. Although kawaii is associated with the general idea of "cute" and often translated into English as "cute," there are nuanced differences between these concepts that psychological research has linked to Japanese culture~\cite{nittono2016two}. Kawaii has positive meanings, such as lovable, adorable, pretty, charming, and sometimes pity~\cite{1,nittono2016two,148,158,159,172}. However, kawaii is not only a matter of appearance, but also a positive affective response that, among Japanese people, can be heightened by the social context~\cite{nittono2022psychology}.

The concept of kawaii has been explored in psychology from cognitive science and behavioural perspectives~\cite{nittono2010behavioral,nittono2016two,nittono2022psychology}. A small amount of work in human-computer interaction (HCI), human-robot interaction (HRI), and human-agent interaction (HAI)~\cite{192,seaborncanvoicesoundcute} has followed suit. A wide array of often disconnected work exists. Kawaii has been explored in agent design (e.g., crafting agents with kawaii features to improve user ratings~\cite{90,142}), user experience or UX (e.g., people having more tolerance~\cite{2}, care responses to agents~\cite{52}), interpretive frameworks (e.g., a visual framework of kawaii attributes~\cite{173}, and user models, including those grounded in participant data~\cite{148,157}).
However, little research has distinguished kawaii and cute clearly in HCI context. This makes it difficult to clarify what is meant by "kawaii," especially when the term is used outside of the Japanese context, i.e., without Japanese researchers, designers, and/or users (e.g., compare \cite{zhang2021exploring}) and \cite{90}). Indeed, there is yet no kawaii computing "field of study" nor consensus, such as by way of a systematic review. This creates a disconnect between HCI and related domains with the strong theoretical and empirical foundation established in psychology~\cite{nittono2010behavioral,nittono2022psychology, nittono2016two}. We appear to have a gap in our understanding of how kawaii has been, could be, and perhaps should be explored in HCI and adjacent fields.

In this preliminary work, we offer a scoping review~\cite{Munn2018,Tricco2018} 
to better understand how kawaii has been used in HCI and related fields as a factor of interaction design and UX. Our goal was to map out and delineate \textbf{kawaii computing} as a concrete domain in HCI. To this end, we asked several sub-questions so as to be comprehensive in our coverage of the research content offered by this wide-reaching literature: What is "kawaii"? How has it been defined and operationalized? What fields and contexts of use has kawaii been explored in? What agents, interfaces, and systems have been deemed to embody kawaii design? What methods have been employed to study it as a UX phenomenon? What factors have been found to influence user perceptions of and reactions to kawaii? How has kawaii been approached as an object of study in HCI?
We contribute a map of research that may be characterized as kawaii computing, as well as identify gaps and opportunities for future work. We set the stage for a more robust kawaii future in HCI.

\section{Methods}

We carried out a scoping review of the literature using the guidelines by \citet{Munn2018} and the protocol offered by the Preferred Reporting Items for Systematic Reviews and Meta-Analyses Extension for Scoping Reviews (PRISMA-ScR)~\cite{Tricco2018}. The purpose of a scoping review, compared to other types of reviews, is to map out the ``scope'' of a topic or domain of study to identify gaps and generate new research directions~\cite{Munn2018}. Typically, scoping reviews reveal the state of affairs and help determine whether a systematic review is possible or necessary. Scoping reviews can also act as a way of solidifying an emerging area of focus, which we believe to be true in the case of kawaii computing. Our process is presented in Supplementary Materials Figure 1. Notably, we deviated from the PRISMA-ScR by not implementing those parts related to conventions in the medical field, e.g., structured abstracts. Our protocol was registered on January 11\textsuperscript{th}, 2024\footnote{\url{https://osf.io/g5knm}}. We provide our data set as part of the Open Science initiative~\cite{natureopenscience2017} here:

\noindent \url{https://bit.ly/kawaiicomputing}.

\subsection{Terms and Definitions}

\subsubsection{Kawaii} We restricted our focus to the Japanese conceptualization of ``kawaii.'' However, reports written in English by and about Japanese people may translate kawaii as ``cute''~\cite{Nittono2012}. We structured our search queries around this possibility by considering ``cute'' within the context of Japan or Japanese participants. During screening, we carefully checked whether the authors or participants were Japanese when ``cute'' rather than ``kawaii'' was used.

\subsubsection{Kawaii Computing} We characterize ``kawaii computing'' as \emph{all human-centred research on kawaii involving computer-based agents, interfaces, environments, and systems}. This broad definition includes virtual characters, UI design patterns, machine learning techniques to process human data, user reactions to chatbots, and so on. The guiding principle is of kawaii as a fundamental psychological phenomenon premised in Japanese culture and experienced as emotion, cognitions, and/or behaviours by Japanese people, in line with the work of Nittono and colleagues~\cite{Nittono2012}.

\subsection{Eligibility Criteria}

We included research reports (short and long papers, conference proceedings and journals, etc.) on kawaii computing (according to the Japanese notion or ``cute'') from 2000 to 2023. We excluded non-human subjects research, researching not involving computer-based systems (not relevant to HCI), grey literature (unverifiable quality or publication status), and papers not in English, Japanese or Chinese languages (the languages known to the authors).

\subsection{Information Sources and Search Strategy}

We searched disciplinary databases relevant to HCI---the ACM Digital Library and IEEE Xplore--as well as Web of Science, a general database to which we had access through our institution, for HCI research or edge cases published outside of ACM and IEEE. We ran our searches between January 3\textsuperscript{rd} and 25\textsuperscript{th}, 2024. Our basic query structure was: \emph{kawaii OR (cute AND Japan) AND (human factors computing OR human-computer interaction OR interaction design)}. Full details for each database search query are presented in \autoref{tab:queries}. We used Zotero to store, remove duplicates, and count the downloaded records returned by these searches. Refer to Supplementary Materials Figure 1 for counts at each stage.

\begin{table*}[h]
\small
\caption{Search queries used in each data base and number of records returned.}
\label{tab:queries}
\begin{tabular}{p{0.09\linewidth}p{0.50\linewidth}p{0.18\linewidth}p{0.05\linewidth}p{0.06\linewidth}}
\toprule
\textit{Database} & \textit{Query} & \textit{Filters} & \textit{Count} & \textit{Date} \\
\midrule
ACM DL & {[}{[}All: "kawaii"{]} OR {[}All: "cute in japan"{]}{]} AND {[}{[}All: "computer"{]} OR {[}All: "computing"{]} OR {[}All: "interactive design"{]} OR {[}ALL:"robot"{]}{]} & / & 43 & 2024.1.3 \\
 & {[}{[}All: "kawaii"{]} OR {[}{[}All: "cute"{]} AND {[}All: "japan"{]}{]}{]} AND {[}{[}All: "human factors computing"{]} OR {[}All: "human-computer interaction"{]} OR {[}All: "interaction design"{]}{]} & Research Article & 146 & 2024.1.10 \\
  & {[}{[}All: "kawaii"{]} OR {[}{[}All: "cute"{]} AND {[}All: "japan"{]}{]}{]} AND {[}{[}All: "human factors computing"{]} OR {[}All: "human-computer interaction"{]} OR {[}All: "interaction design"{]}{]} & Abstract, Poster, Extended Abstract, Short paper, Work in progress & 38 & 2024.1.24 \\
IEEE Xplore & ((kawaii OR (cute AND Japan)) AND (human factors computing OR human-computer interaction OR interaction design) ) & / & 10 & 2024.1.10 \\
\makecell[lt]{Web of \\Science} & ((kawaii OR (cute AND Japan)) AND (human factors computing OR human-computer interaction OR interaction design) )(All Fields) & / & 44 & 2024.1.10\\
\bottomrule
\end{tabular}
\end{table*}

\subsection{Screening and Conflict Management}

Two authors carried out two screening stages.
The first involved checking the title, abstract, and keywords for face viability. This was done by the first author. The second author then checked the excluded records, re-including as needed. The second phase of screening involved both reading pertinent parts of the full texts with the aid of keyword searches to better determine viability. Importantly, we expected to find one of the author's own work returned. When this occurred, screening, data extraction and synthesis of these studies was conducted by the other author.

\subsection{Data Items and Data Extraction}

We used Google Sheets to screen, extract data, and carry out analysis of the data. Specifically, we extracted the following: metadata, including Authors, Year of publication, Title, Abstract, and URL; Topic, specifically Topic of study	and Context; Terms, including Definition of kawaii, whether Explicit/implied, Citations for definitions/, and whether there was No mention of kawaii but from the Japanese context; Stimuli, including Stimuli details, the Measure of kawaii, and what Kawaii factors were identified; Methods, including Type of research, Research design, Participant nationality, Demographics, Sample size, and Data Analysis; Agent Types, including Virtual agents, Robots with a physical body,	Interfaces, and Other; and How was Kawaii used, including Design (Feature, Control), User Reaction (Self-Reports, Biometrics), and Interpretive Framework (Data Analysis, Machine Learning, Heuristic analysis).

\subsection{Data Analysis}

We carried out two data analyses. For quantitative data, we generated descriptive statistics, including counts and percentages. For qualitative data, the first author used content analysis to categorize the "broad surface structure," focusing on surface meanings rather than latent interpretations, also called a "manifest analysis"~\cite{Bengtsson2016}.

\section{Results}

We identified 238 records through database searching. After screening and eligibility, 69 records were included (Supplementary Materials Figure 1). The full list of included papers is available at \url{https://bit.ly/kawaiicomputing}


\subsection{Kawaii in Topics and Contexts}

We found seven contexts and a variety of topics in kawaii computing. The most common context was HRI (n=29, 42\%), followed by HCI (n=16, 23\%) and HAI (n=15, 22\%). Several papers covered machine learning (ML) (n=4, 6\%), human-machine interaction (HMI) (n=2, 3\%), culture (n=2, 3\%) and human-human communication (n=1, 1\%).

Although covering a wide array of topics, most papers only referenced kawaii in participants' open-ended answers and the authors' discussion. Less than half (n=31, 45\%) included kawaii as a main topic. 
In HAI, models for kawaii virtual agents were proposed, e.g., \citet{162} set up kawaii attributes for evaluating kawaii virtual characters and \citet{1} proposed a preliminary model for kawaii game vocalics. Others evaluated user feelings of kawaii-ness towards virtual characters~\cite{159,176}, 3D models of animals~\cite{150}, VTubers~\cite{2}, and preferences for kawaii and non-kawaii game characters~\cite{63}. In HRI, kawaii was framed as a feature of robot design, with the focus on identifying kawaii factors, e.g., NAO robot \cite{172}, 3D robots \cite{160,161}, commercial delivery robots \cite{192}, virtual cross-cultural kawaii robots \cite{158}, and the design of Pepper's touch motion to express kawaii~\cite{149}. In HCI, kawaii was applied in social media, e.g., \citet{174} proposed a kawaii search engine for blogs, and \citet{166} proposed and evaluated LINE's kawaii chat interface. Kawaii was also introduced to design other interfaces, such as the appearance of Kissenger (a remote communication device) \cite{90}, the LiveDeck interface for real-time collaborative editing \cite{142}, and a research through design (RtD) critical analysis of the aesthetic object design fiction "Menstruation Machine" \cite{140}. In ML,  \citet{173} and \citet{160} developed framework to classify kawaii fashion with colour and clothing. \citet{148} also proposed a model to compare the kawaiiness of online images of cosmetic bottles. \citet{39} explored the influence of culture-specific narratives on the imagination of future AI technology, in which kawaii is viewed as a specific feature of Japanese culture.

\subsection{Distribution of Kawaii Computing Research}

Since kawaii is a concept from Japan and directly tied to Japanese culture,
we considered participant demographics. 
Most studies included Japanese participants (n=56, 81\%). Among these, 44 papers included only Japanese participants (64\%) and 12 included both Japanese and participants from other countries (17\%) and aimed at cross-cultural comparison or generalization. 
For example, \citet{161} compared kawaii robot designs by Japanese and U.S. students, and \citet{148,159} included both Japanese and Thai participants to establish a framework for kawaii feeling evaluations. Some papers did not include Japanese participants (7\%). These papers introduced "kawaii" into other culture contexts. For example, \citet{2} studied kawaii as a part of otaku culture, which influenced Chinese viewer perceptions and attitudes towards live-stream VTubers. \citet{172} also had a study in Chinese context to investigate if kawaii design is valuable to Chinese customers. The three papers about ML had no participants (6\%). Other demographics used were gender, age, occupation, education, etc. Notably, most work included combinations of women and men, except one paper for a new HMI interface for a straddle-type vehicle design tested with men~\cite{70}. One paper investigated the effect of robotic mediation in promoting conversation among older adults for improving behavioural and psychological symptoms of dementia (BPSD) and included only women \cite{66}. Another was about a system with kawaii (cute in Japan) icons for product retrieval and only tested women because the design was inspired by the psychology of women's shopping activity \cite{122}. Many papers (n=17, 25\%) focused on students, from junior high to college, and another six (9\%) focused on younger generations (under 30 years old) \cite{81,90,130,208,218,219}. Four focused on kawaii computing for older adults (7\%), e.g., comparing perceptions of kawaii between the young and the old \cite{172}, a kawaii social robot for children-older adult interaction \cite{52}, and conversational robots for a care center \cite{65} and older adults with dementia \cite{66}. Two (4\%) conducted research on children: a kawaii social robot mediating children-older adult interaction \cite{52}, and delivering negative feedback for improving the in-class learning experience \cite{80}.

\subsection{Definition of Kawaii}

More than a half papers (n=41, 65\%) did not mention the word "kawaii," but we can infer that the use of "cute" meant "kawaii" because of the Japanese context. We equate uses of "cute" within the Japanese context to the concept of kawaii. In 96 papers, 24 (35\%) used the definition of kawaii or cute in the Japanese context. 17 (25\%) provided an explicit definition (13 with citations, 4 without citations), and nine (13\%) provided an implicit definition (5 with citations, 4 without citations). The citations were sourced from critical, cultural, and psychological studies.

For explicit definitions, most wrote that kawaii means Japanese cuteness, period. 
\citet[abstract]{161} defined kawaii as "a Japanese adjective representing cute and adorable" \cite{marcus2017cuteness,ohkura2019kawaii,Nittono2012,nittono2019kawaii}.
\citet{1,234} characterized kawaii as the Japanese concept of "cute++," linked to cognition, behaviour, and emotion, citing \citet{nittono2010behavioral}. They also referenced impressions of cuteness, charm, endearment, and pity.
Similarly, \citet[p. 1126]{142} explained that kawaii is "an aspect of Japanese culture that refers to cute things that evoke positive emotions and feelings of social affiliation," also citing \citet{nittono2010behavioral}.
Some mentioned that kawaii has positive connotations.
\citet{158} used the same definition in their previous work \cite{166} that described kawaii as Japanese cuteness, a concept ingrained in Japanese contemporary society. They looked into Japanese history and literature, which described kawaii as pitiful, shameful, or too sad to see, small, weak, and someone or something that gives one the feeling of “wanting to protect” \cite{yomota2006kawaii}.
\citet{148,159,160} defined kawaii as a positive adjective that denotes such meanings as cute or lovable. They state that it is critical as a kansei value plays a leading role in the worldwide success of Japanese products, such as Hello Kitty \cite{kovarovic2011hello} and Pokémon \cite{allison2003portable}.
\citet{149} and \citet{205} defined kawaii as a Japanese word that has positive connotations as cute \cite{Nittono2012,nittono2016two}, which is employed when designing robot. \citet[p. 11276]{149} position kawaii as is a critical factor in Japanese commercial aspects and pop culture \cite{kinsella2013cuties,lieber2019word}.

Others provided a definition of kawaii in design. \citet{174} linked kawaii with pretty or lovely design.
\citet[258]{172} characterized kawaii as "an affective concept rooted in Japanese aesthetics". For artificial products, they thought kawaii is a critical factor of Japanese kansei design, which can mean cute, lovely, small, and other related emotional values \cite{ohkura2010systematic}.
\citet{192} described kawaii in the context of robot design. Kawaii, an important concept in Japanese pop culture and design \cite{149}, has been operationalized in some studies in HRI \cite{berque2020fostering}, as not only the perceivable features of robots, but also sociality and approach motivation \cite{192,nittono2022psychology}. \citet{230} also described kawaii in robot design as descriptive features that characterize a robot, such as manga eyes and low resemblance to a human.
Some related kawaii with girlish. \citet[p.  4]{176} proposed that "kawaii has become an important component of aesthetics in Japanese consumer culture and has also spread to international markets and audiences", which became a popular way to describe female idols in 80s Japanese pop culture \cite{sone2016japanese}. They also argued that kawaii inspired the performance of cuteness, innocence, and femininity which contributes to the fetishization and commodification of a particular type of culturally-desired gender dynamics: cute artificial women and male consumers \cite{black2008virtual}.
\citet{140} described kawaii style as Japanese cute girl aesthetic characterized by rainbow-colored hair, glamour make-up and so on.

Implicit definitions of kawaii were introduced with reference to other studies or general knowledge, without author clarification. Kawaii was described as am important part of otaku culture \cite{2}, by way of agents like Kewpie \cite{39}, which is one of the most universally recognized symbols of cuteness in Japan \cite{riessland1998sweet}, attractive in a pretty or endearing way \cite{211}, an appearance similar to a stuffed toy \cite{90,213}, borderline before reaching the Uncanny Valley \cite{cheok2010kawaii}, and Harajuku street fashion as kawaii fashion \cite{173}. \citet{162} and \citet{173} agreed that kawaii attracts attention as an affective value. \citet{173} further mentioned there are various kawaii attributes for design, such as color \cite{ohkura2014kawaii}.

We used word frequency analysis to analyze the 20 definitions extracted. The top five frequency words (except kawaii) were "Japanese" (n=27), "cute" (n=13), "cuteness" (n=9), "culture" (n=8), "positive" (n=6), "pop" (n=5). "Japan," "affective," "products," and "important" (n=4) followed. This indicates that the definition of kawaii is highly related to a Japanese pop and consumer culture, is deemed connected to "cute," and has positive meanings.

\subsection{Kawaii Stimuli, Measurement, and Factors}

We found six general stimuli: visual, audio, movements, gestures, and text. Most of studies focus on one stimuli (n=49, 71\%). The most common stimuli was visual (n=83, 87\%), followed by movements (n=12, 17\%), audio (n=8, 12\%), gestures (n=2, 3\%), text (n=1, 1\%), and animation (n=1, 1\%). The specific stimuli for visual were various, e.g., virtual agents, robots, avatars, images, emotes, etc. The movement stimuli were mainly robots with entities (n=8) followed by interfaces (n=2) \cite{208,213}, virtual agents (n=2) \cite{81,232}, and cars with eyes \cite{232}. As for audio stimuli, the specific stimuli were game vocalics \cite{1}, robot \cite{82,120,134}, Hatsune Miku virtual character and TTS voice \cite{176}, and mucis vidio \cite{140}. Specific stimuli of gestures were virtual agent \cite{18}, and robots \cite{58}.

41 papers (59\%) proposed measurements of kawaii 
(refer to Supplementary Table 1 for full details). Among them, six papers measured kawaiiness without discussing kawaii factors. On the other hand, 15 papers discussed kawaii factors but did not provide or use measurements of kawaii. 
Researchers considered a variety of kawaii factors, but given the variety of the research, it is hard to summarize, so we describe the kawaii factors that occurred most frequently in our corpus. 
These were: physical features related to Baby Schema; anthropomorphism, which is an important feature for non-living things; and factors related to Japanese culture, such as animation, girlishness, and colour. \citet{6,44,63,99,158,161} deemed the physical features of size and shape as one of the most common factors, which is consistent with Baby Schema. They found that a round shape~\cite{99,158} and small size~\cite{6,63} was perceived as more kawaii, because people felt less physical pressure or stress~\cite{44}. Anthropomorphism was another common factor in robot kawaiiness for visual but also audio stimuli~\cite{1,94}. Human-like robots, animal-like robots, and virtual agents~\cite{80, 91} were deemed most kawaii. Animation and manga styling, which is a symbol of Japanese pop culture, were linked to greater levels of kawaii \cite{205,230}. Femininity is another common factor related to high levels of kawaiiness, especially girlishness~\cite{1,162,176}. Colour was a common kawaii factor. \citet{99} found that people perceived colourful drones as more kawaii. \citet{157,158,161,172} found similar colour factors and results linked to kawaiiness.

\subsection{Use of Kawaii}

Kawaii has been used in different ways in HCI and related areas. 
We summarized how it has been approached in~\autoref{tab:summary}. Kawaii has be applied in HCI research for design, UX measurement, and establishment of an interpretive framework. In HCI design, kawaii has been used as a feature that makes robots more acceptable to people, especially social robots and interactive communication agents. Instead of embedding kawaii features in design, \citet{150} controlled different levels of kawaii and compared these in 3D models to find a better design. Another study directly included the meaning of kawaii using the Japanese word "Puyo" in the product name to appeal to users~\cite{225}. Kawaii, which is related to emotions, can act as a user reaction for UX evaluations. Furthermore, kawaii has been included in interpretive frameworks. \citet{159} constructed a mathematical model of kawaii feelings, while \citet{148} proposed a machine learning model to evaluate kawaii attributes of kawaii products.  \citet{176} used heuristic analysis to study the kawaii impacts on virtual agent design. Kawaii thus has versatility as a feature of design and a measurable user reaction or experience.

\begin{table*}[!ht]
    \centering
      \caption{Summary of how kawaii has been approached in HCI and adjacent domains.}
    \label{tab:summary}
    \begin{tabular}{p{0.1\linewidth}p{0.2\linewidth}p{0.51\linewidth}p{0.095\linewidth}}
    \toprule
    Components&Details&Papers&Count (\%)\\
    \midrule
    Design&Feature&\cite{18,39,52,57,82,90,98,99,134,138,142,148,157,158,159,160,161,162,166,172,173,176,192,140,234,205,208,211,213,215,218,220,230}&34 (49\%)\\
    &Control&\cite{150}&1 (1\%)\\
    &Name&\cite{225}&1 (1\%)\\
    
    \midrule
    \makecell[lt]{User\\Reaction}&Self-reports&\cite{1,2,6,8,10,30,32,41,58,61,63,64,65,66,70,71,80,81,83,91,92,94,99,120,122,130,134,142,149,150,158,159,160,161,162,172,174,192,140,234,196,202,208,210,215,218,219,220,230,232}&50 (72\%)\\
    &Biometrics&\cite{159,161}&2 (3\%)\\
    
    \midrule
    \multirow{2}{*}{\makecell[lt]{Interpretive\\Framework}}
    &Data Analysis&\cite{159}&1 (1\%)\\
    &Machine Learning (ML)&\cite{39,109,148,157,162,173}&6 (9\%)\\
    &Heuristic Analysis&\cite{174,176}&2 (3\%)\\
    \bottomrule
    \end{tabular}
\end{table*}

\section{Discussion}

With the global spread of kawaii and Japanese culture, attention to the notion of "cute" as a cultural facet of Japan is on the rise. Kawaii research and design practice is also steadily increasing. When it comes to HCI and related fields, how do we ride this trend? Although existing studies were not exhaustive, they do provide inspiration and directions for future work.

\subsection{Kawaii vs. Cute}

The review of kawaii allows us to have an understanding of the differences between kawaii and cute. Compared to cute, kawaii embodies a particular Japanese aesthetic and culture frame \cite{2,142,157,166,176}. It goes beyond just being cute, as cute does not have any cultural connotations. Cute works in HCI typically applies the "Baby Schema" concept defined by Lorenz in 1943 \cite{lorenz1943innate} to guide the creation of interactive artefacts \cite{fan2023cuteness}.
Kawaii, however, is not limited to being a Baby Schema phenomenon, nor is it simply a feature of design \cite{nittono2019kawaii}.
\citet{nittono2016two} proposed a two-layer model of kawaii, with one layer as emotion and the other layer as social value. The kawaii as an emotion frame presents kawaii as a stimulus perceived as cute, friendly, harmless, pretty and so on, regardless of whether the stimulus is a living being. When kawaii affect is evaluated as significant through cognitive appraisal processes, it can be linked to a psychological state of kawaii \cite{nittono2016two}. Kawaii is also framed as social value, referring to kawaii as a medium of sociality \cite{nittono2016two}.
This has be adopted in many kawaii works as well as in HCI (e.g., \cite{1,142,149,158,161}). A more latent feature of kawaii is its basis in Japanese culture. \citet{nittono2019kawaii} notes that features particular to Japanese culture appear to influence and temper the concept of kawaii, such as amae, one's drive for acceptance and adoration from other people, and chizimi shikou, the love of small things. We found no specific reference to these yet.

Research in kawaii computing is not the same topically as research on cute design based on Baby Schema in HCI.
Kawaii is not only the feature of the design of a robot, virtual agent, or interface, but also emotion that occurs in the interaction between human-agent, human-robot, and human-interface (and perhaps between people with a facilitating computer). We propose that kawaii computing provides an approach to improve the relationship between people and technologies. Moreover, kawaii may vary in different social contexts, especially different culture. For example, a robot may become kawaii when it interacts with your pet \cite{120}. Although work is limited (17\%), some cross-cultural research has found that Japanese people are more sensitive and connected to kawaii computing \cite{39,142,158,159}. As expected, most kawaii computing work was done by Japanese researchers and included Japanese subjects (81\%). Still, kawaii as a psychological state of feeling positive emotions \cite{nittono2016two} is not limited in Japan. This shows a gap in cross-cultural kawaii design and user perceptions of kawaii towards robots, agents, interfaces, and so on.

\subsection{An Emerging Map of Kawaii Computing}
Kawaii was found across HCI and adjacent contexts, with work in design (51\%), user reactions (75\%), and interpretive frameworks (13\%).
This indicates that kawaii can play several different roles in research and design. 
Most work studied user reactions through participants' kawaii emotions. There was limited work on kawaii data analysis frameworks \cite{159}, ML structures for kawaii evaluation \cite{157,162,173}, and heuristic analysis \cite{173,174}. However, even much work covered kawaii user reactions, there was no universal framework for kawaii measurement. Most researchers created their own scales \cite{1,8,70,81,162,220,230}, some employed (un/validated) items from other kawaii works \cite{nittono2019kawaii, nittono2016two} in psychological and behavioural fields \cite{149,150,218}, and two used the objective measurement of biometrics \cite{161,172}.
Many (28\%) studied kawaii both as a feature of design and as a user reaction, which is consistent with the two-factor model of kawaii: not only features but also kawaii emotions in the interaction.
Kawaii attributes of design were various within and across agents, robots, and interfaces~\cite{94,99,157,205,230,232}, which is similar to the situation of kawaii measures. This shows a need for researchers to establish a validated frameworks for understanding the components of kawaii computing.

\subsection{An Agenda for Future Work}
Based on this scoping review, we offer some ideas and prompts for future work on kawaii computing.

\begin{itemize}
 \item \textbf{Standardization: Devise a taxonomy of attributes (stimulus design) and standard measures (user response) for kawaii computing.} In HCI and related domains, kawaii attributes across agents, robots, and interface designs as well as measures of evaluating kawaii are fragmented. Most relied on (unvalidated) self-designed materials based on the specific context. Establishing sets of attributes for kawaii computing and interpretive frameworks for evaluating kawaii reactions in HCI, HRI, and HAI will be a key endeavour.
 \item \textbf{Multiple Modalities: Explore kawaii computing with stimuli beyond the visual.} The body of work covered in this review indicates a focus on visual stimuli. While only a few works involved other stimuli, such as audio \cite{1,234}, HCI is not always visually based, and sometimes uses only non-visual modalities, e.g., voice assistants. 
  \item \textbf{Cultural Significance: Adapt kawaii cross-culturally.} 
  "Kawaii" is defined by and has a reciprocal relationship with Japanese culture. Most research has been conducted in Japan or by Japanese people. However, the power of kawaii may not only be limited to the Japanese context. Future work can study the cross-cultural effects on and differences in kawaii perceptions and experiences. For example, comparisons of Japanese and non-Japanese foreigners living in Japan, or comparisons of kawaii computing artefacts and computing artefacts designed to be cute in non-Japanese contexts.
 \item \textbf{Sociality: Evaluate the social impacts of kawaii computing platforms.} 
 Sociality refers to social interactions between users and agents and social reactions to kawaii stimuli. We may also need to explore the impact on society when it comes to kawaii computing platforms deployed at scale. 
 Future work should explore how the social plays out so as to understand the social impacts of kawaii computing.
 \item \textbf{Adjacent Concepts: Explore the theorized relevance of other aspects of Japanese culture connected to kawaii, such as amae and chizimi shikou.} Amae may also relate to sociality. Could kawaii chatbots make us feel loved and accepted? Chizimi shikou could help explain preferences to small devices and robots.
\end{itemize}

\section{Conclusion}

Kawaii computing is nascent. While not exactly new, the lack of formalization around what it is leaves this area of study brimming with possibility. Researchers and designers across HCI, HRI, HAI, and beyond should draw from the formal theories and models established in psychology. We have the opportunity now to build on this foundation and direct the future of kawaii computing in a more rigorous, culturally-sensitive, and theoretically-driven way.

\begin{acks}
Yijia Wang is grateful to the China Government Scholarship that sponsored this work as part of her doctoral programme. Katie Seaborn conscientiously dissents to in-person participation at CHI; read their positionality statement here: \url{https://bit.ly/chi24statement}
\end{acks}

\bibliographystyle{ACM-Reference-Format}
\bibliography{bib}


\begin{thebibliography}{96}


\ifx \showCODEN    \undefined \def \showCODEN     #1{\unskip}     \fi
\ifx \showDOI      \undefined \def \showDOI       #1{#1}\fi
\ifx \showISBNx    \undefined \def \showISBNx     #1{\unskip}     \fi
\ifx \showISBNxiii \undefined \def \showISBNxiii  #1{\unskip}     \fi
\ifx \showISSN     \undefined \def \showISSN      #1{\unskip}     \fi
\ifx \showLCCN     \undefined \def \showLCCN      #1{\unskip}     \fi
\ifx \shownote     \undefined \def \shownote      #1{#1}          \fi
\ifx \showarticletitle \undefined \def \showarticletitle #1{#1}   \fi
\ifx \showURL      \undefined \def \showURL       {\relax}        \fi
\providecommand\bibfield[2]{#2}
\providecommand\bibinfo[2]{#2}
\providecommand\natexlab[1]{#1}
\providecommand\showeprint[2][]{arXiv:#2}

\bibitem[Allen et~al\mbox{.}(2012)]%
        {41}
\bibfield{author}{\bibinfo{person}{Jeffrey Allen}, \bibinfo{person}{James~E. Young}, \bibinfo{person}{Daisuke Sakamoto}, {and} \bibinfo{person}{Takeo Igarashi}.} \bibinfo{year}{2012}\natexlab{}.
\newblock \showarticletitle{Style by demonstration for interactive robot motion}. In \bibinfo{booktitle}{\emph{Proceedings of the Designing Interactive Systems Conference}} (Newcastle Upon Tyne, United Kingdom) \emph{(\bibinfo{series}{DIS '12})}. \bibinfo{publisher}{Association for Computing Machinery}, \bibinfo{address}{New York, NY, USA}, \bibinfo{pages}{592–601}.
\newblock
\showISBNx{9781450312103}
\urldef\tempurl%
\url{https://doi.org/10.1145/2317956.2318045}
\showDOI{\tempurl}


\bibitem[Allison(2003)]%
        {allison2003portable}
\bibfield{author}{\bibinfo{person}{Anne Allison}.} \bibinfo{year}{2003}\natexlab{}.
\newblock \showarticletitle{Portable monsters and commodity cuteness: Poke{\'{}} mon as Japan's new global power}.
\newblock \bibinfo{journal}{\emph{Postcolonial Studies}} \bibinfo{volume}{6}, \bibinfo{number}{3} (\bibinfo{year}{2003}), \bibinfo{pages}{381--395}.
\newblock


\bibitem[Bardzell et~al\mbox{.}(2015)]%
        {140}
\bibfield{author}{\bibinfo{person}{Jeffrey Bardzell}, \bibinfo{person}{Shaowen Bardzell}, {and} \bibinfo{person}{Lone Koefoed~Hansen}.} \bibinfo{year}{2015}\natexlab{}.
\newblock \showarticletitle{Immodest Proposals: Research Through Design and Knowledge}. In \bibinfo{booktitle}{\emph{Proceedings of the 33rd Annual ACM Conference on Human Factors in Computing Systems}} (Seoul, Republic of Korea) \emph{(\bibinfo{series}{CHI '15})}. \bibinfo{publisher}{Association for Computing Machinery}, \bibinfo{address}{New York, NY, USA}, \bibinfo{pages}{2093–2102}.
\newblock
\showISBNx{9781450331456}
\urldef\tempurl%
\url{https://doi.org/10.1145/2702123.2702400}
\showDOI{\tempurl}


\bibitem[Bengtsson(2016)]%
        {Bengtsson2016}
\bibfield{author}{\bibinfo{person}{Mariette Bengtsson}.} \bibinfo{year}{2016}\natexlab{}.
\newblock \showarticletitle{How to plan and perform a qualitative study using content analysis}.
\newblock \bibinfo{journal}{\emph{NursingPlus Open}}  \bibinfo{volume}{2} (\bibinfo{year}{2016}), \bibinfo{pages}{8–14}.
\newblock
\showISSN{2352-9008}
\urldef\tempurl%
\url{https://doi.org/10.1016/j.npls.2016.01.001}
\showDOI{\tempurl}


\bibitem[Berque and Chiba(2017)]%
        {166}
\bibfield{author}{\bibinfo{person}{Dave Berque} {and} \bibinfo{person}{Hiroko Chiba}.} \bibinfo{year}{2017}\natexlab{}.
\newblock \showarticletitle{Evaluating the Use of LINE Software to Support Interaction During an American Travel Course in Japan}. In \bibinfo{booktitle}{\emph{Lecture Notes in Computer Science (including subseries Lecture Notes in Artificial Intelligence and Lecture Notes in Bioinformatics)}}. \bibinfo{publisher}{Springer Nature}, \bibinfo{address}{London, UK}, \bibinfo{pages}{614--623}.
\newblock
\showISBNx{978-3-319-57930-6}
\urldef\tempurl%
\url{https://doi.org/10.1007/978-3-319-57931-3_49}
\showDOI{\tempurl}


\bibitem[Berque et~al\mbox{.}(2022)]%
        {158}
\bibfield{author}{\bibinfo{person}{Dave Berque}, \bibinfo{person}{Hiroko Chiba}, \bibinfo{person}{Tipporn Laohakangvalvit}, \bibinfo{person}{Michiko Ohkura}, \bibinfo{person}{Peeraya Sripian}, \bibinfo{person}{Midori Sugaya}, \bibinfo{person}{Liam Guinee}, \bibinfo{person}{Shun Imura}, \bibinfo{person}{Narumon Jadram}, \bibinfo{person}{Rafael Martinez}, \bibinfo{person}{Sheong~Fong Ng}, \bibinfo{person}{Haley Schwipps}, \bibinfo{person}{Shuma Ohtsuka}, {and} \bibinfo{person}{Grace Todd}.} \bibinfo{year}{2022}\natexlab{}.
\newblock \showarticletitle{Cross-Cultural Design and Evaluation of Student Companion Robots with Varied Kawaii (Cute) Attributes}. In \bibinfo{booktitle}{\emph{Human-Computer Interaction. Theoretical Approaches and Design Methods: Thematic Area, HCI 2022, Held as Part of the 24th HCI International Conference, HCII 2022, Virtual Event, June 26–July 1, 2022, Proceedings, Part I}}. \bibinfo{publisher}{Springer-Verlag}, \bibinfo{address}{Berlin, Heidelberg}, \bibinfo{pages}{391–409}.
\newblock
\showISBNx{978-3-031-05310-8}
\urldef\tempurl%
\url{https://doi.org/10.1007/978-3-031-05311-5_27}
\showDOI{\tempurl}


\bibitem[Berque et~al\mbox{.}(2020)]%
        {berque2020fostering}
\bibfield{author}{\bibinfo{person}{Dave Berque}, \bibinfo{person}{Hiroko Chiba}, \bibinfo{person}{Michiko Ohkura}, \bibinfo{person}{Peeraya Sripian}, {and} \bibinfo{person}{Midori Sugaya}.} \bibinfo{year}{2020}\natexlab{}.
\newblock \showarticletitle{Fostering cross-cultural research by cross-cultural student teams: A case study related to kawaii (Cute) robot design}. In \bibinfo{booktitle}{\emph{Cross-Cultural Design. User Experience of Products, Services, and Intelligent Environments: 12th International Conference, CCD 2020, Held as Part of the 22nd HCI International Conference, HCII 2020, Copenhagen, Denmark, July 19--24, 2020, Proceedings, Part I 22}}. \bibinfo{publisher}{Springer}, \bibinfo{address}{London, UK}, \bibinfo{pages}{553--563}.
\newblock


\bibitem[Black(2008)]%
        {black2008virtual}
\bibfield{author}{\bibinfo{person}{Daniel Black}.} \bibinfo{year}{2008}\natexlab{}.
\newblock \showarticletitle{The virtual ideal: Virtual idols, cute technology and unclean biology}.
\newblock \bibinfo{journal}{\emph{Continuum}} \bibinfo{volume}{22}, \bibinfo{number}{1} (\bibinfo{year}{2008}), \bibinfo{pages}{37--50}.
\newblock


\bibitem[Chang et~al\mbox{.}(2022)]%
        {138}
\bibfield{author}{\bibinfo{person}{Chia-Ming Chang}, \bibinfo{person}{Koki Toda}, \bibinfo{person}{Xinyue Gui}, \bibinfo{person}{Stela~H. Seo}, {and} \bibinfo{person}{Takeo Igarashi}.} \bibinfo{year}{2022}\natexlab{}.
\newblock \showarticletitle{Can Eyes on a Car Reduce Traffic Accidents?}. In \bibinfo{booktitle}{\emph{Proceedings of the 14th International Conference on Automotive User Interfaces and Interactive Vehicular Applications}} (Seoul, Republic of Korea) \emph{(\bibinfo{series}{AutomotiveUI '22})}. \bibinfo{publisher}{Association for Computing Machinery}, \bibinfo{address}{New York, NY, USA}, \bibinfo{pages}{349–359}.
\newblock
\showISBNx{9781450394154}
\urldef\tempurl%
\url{https://doi.org/10.1145/3543174.3546841}
\showDOI{\tempurl}


\bibitem[Chen et~al\mbox{.}(2021)]%
        {220}
\bibfield{author}{\bibinfo{person}{Dominique Chen}, \bibinfo{person}{Young~ah Seong}, \bibinfo{person}{Hiraku Ogura}, \bibinfo{person}{Yuto Mitani}, \bibinfo{person}{Naoto Sekiya}, {and} \bibinfo{person}{Kiichi Moriya}.} \bibinfo{year}{2021}\natexlab{}.
\newblock \showarticletitle{Nukabot: Design of Care for Human-Microbe Relationships}. In \bibinfo{booktitle}{\emph{Extended Abstracts of the 2021 CHI Conference on Human Factors in Computing Systems}} (Yokohama, Japan) \emph{(\bibinfo{series}{CHI EA '21})}. \bibinfo{publisher}{Association for Computing Machinery}, \bibinfo{address}{New York, NY, USA}, Article \bibinfo{articleno}{291}, \bibinfo{numpages}{7}~pages.
\newblock
\showISBNx{9781450380959}
\urldef\tempurl%
\url{https://doi.org/10.1145/3411763.3451605}
\showDOI{\tempurl}


\bibitem[Cheok and Cheok(2010)]%
        {cheok2010kawaii}
\bibfield{author}{\bibinfo{person}{Adrian~David Cheok} {and} \bibinfo{person}{Adrian~David Cheok}.} \bibinfo{year}{2010}\natexlab{}.
\newblock \showarticletitle{Kawaii/cute interactive media}.
\newblock \bibinfo{journal}{\emph{Art and technology of entertainment computing and communication: Advances in interactive new media for entertainment computing}}  \bibinfo{volume}{11} (\bibinfo{year}{2010}), \bibinfo{pages}{223--254}.
\newblock


\bibitem[Cho et~al\mbox{.}(2009)]%
        {61}
\bibfield{author}{\bibinfo{person}{Heeryon Cho}, \bibinfo{person}{Toru Ishida}, \bibinfo{person}{Naomi Yamashita}, \bibinfo{person}{Tomoko Koda}, {and} \bibinfo{person}{Toshiyuki Takasaki}.} \bibinfo{year}{2009}\natexlab{}.
\newblock \showarticletitle{Human detection of cultural differences in pictogram interpretations}. In \bibinfo{booktitle}{\emph{Proceedings of the 2009 International Workshop on Intercultural Collaboration}} (Palo Alto, California, USA) \emph{(\bibinfo{series}{IWIC '09})}. \bibinfo{publisher}{Association for Computing Machinery}, \bibinfo{address}{New York, NY, USA}, \bibinfo{pages}{165–174}.
\newblock
\showISBNx{9781605585024}
\urldef\tempurl%
\url{https://doi.org/10.1145/1499224.1499250}
\showDOI{\tempurl}


\bibitem[Choi et~al\mbox{.}(2019)]%
        {109}
\bibfield{author}{\bibinfo{person}{Saemi Choi}, \bibinfo{person}{Shun Matsumura}, {and} \bibinfo{person}{Kiyoharu Aizawa}.} \bibinfo{year}{2019}\natexlab{}.
\newblock \showarticletitle{Assist Users' Interactions in Font Search with Unexpected but Useful Concepts Generated by Multimodal Learning}. In \bibinfo{booktitle}{\emph{Proceedings of the 2019 on International Conference on Multimedia Retrieval}} (Ottawa ON, Canada) \emph{(\bibinfo{series}{ICMR '19})}. \bibinfo{publisher}{Association for Computing Machinery}, \bibinfo{address}{New York, NY, USA}, \bibinfo{pages}{235–243}.
\newblock
\showISBNx{9781450367653}
\urldef\tempurl%
\url{https://doi.org/10.1145/3323873.3325037}
\showDOI{\tempurl}


\bibitem[Cooney et~al\mbox{.}(2014)]%
        {32}
\bibfield{author}{\bibinfo{person}{Martin Cooney}, \bibinfo{person}{Shuichi Nishio}, {and} \bibinfo{person}{Hiroshi Ishiguro}.} \bibinfo{year}{2014}\natexlab{}.
\newblock \showarticletitle{Affectionate Interaction with a Small Humanoid Robot Capable of Recognizing Social Touch Behavior}.
\newblock \bibinfo{journal}{\emph{ACM Trans. Interact. Intell. Syst.}} \bibinfo{volume}{4}, \bibinfo{number}{4}, Article \bibinfo{articleno}{19} (\bibinfo{date}{dec} \bibinfo{year}{2014}), \bibinfo{numpages}{32}~pages.
\newblock
\showISSN{2160-6455}
\urldef\tempurl%
\url{https://doi.org/10.1145/2685395}
\showDOI{\tempurl}


\bibitem[Dobrosovestnova et~al\mbox{.}(2022)]%
        {192}
\bibfield{author}{\bibinfo{person}{Anna Dobrosovestnova}, \bibinfo{person}{Isabel Schwaninger}, {and} \bibinfo{person}{Astrid Weiss}.} \bibinfo{year}{2022}\natexlab{}.
\newblock \showarticletitle{With a Little Help of Humans. An Exploratory Study of Delivery Robots Stuck in Snow}. In \bibinfo{booktitle}{\emph{2022 31st IEEE International Conference on Robot and Human Interactive Communication (RO-MAN)}}. \bibinfo{publisher}{IEEE}, \bibinfo{address}{New York, NY, USA}, \bibinfo{pages}{1023--1029}.
\newblock
\urldef\tempurl%
\url{https://doi.org/10.1109/RO-MAN53752.2022.9900588}
\showDOI{\tempurl}


\bibitem[Edirisinghe et~al\mbox{.}(2023)]%
        {82}
\bibfield{author}{\bibinfo{person}{Sachi Edirisinghe}, \bibinfo{person}{Satoru Satake}, {and} \bibinfo{person}{Takayuki Kanda}.} \bibinfo{year}{2023}\natexlab{}.
\newblock \showarticletitle{Field Trial of a Shopworker Robot with Friendly Guidance and Appropriate Admonishments}.
\newblock \bibinfo{journal}{\emph{J. Hum.-Robot Interact.}} \bibinfo{volume}{12}, \bibinfo{number}{3}, Article \bibinfo{articleno}{34} (\bibinfo{date}{apr} \bibinfo{year}{2023}), \bibinfo{numpages}{37}~pages.
\newblock
\urldef\tempurl%
\url{https://doi.org/10.1145/3575805}
\showDOI{\tempurl}


\bibitem[Fan et~al\mbox{.}(2023)]%
        {fan2023cuteness}
\bibfield{author}{\bibinfo{person}{Angela~YH Fan}, \bibinfo{person}{Chen Ji}, \bibinfo{person}{Ella Dagan}, \bibinfo{person}{Samir Ghosh}, \bibinfo{person}{Yuhui Wang}, {and} \bibinfo{person}{Katherine Isbister}.} \bibinfo{year}{2023}\natexlab{}.
\newblock \showarticletitle{The Cuteness Factor: An Interpretive Framework for Artists, Designers and Engineers}. In \bibinfo{booktitle}{\emph{Proceedings of the 2023 ACM Designing Interactive Systems Conference}}. \bibinfo{publisher}{ACM}, \bibinfo{address}{New York, NY, USA}, \bibinfo{pages}{2509--2521}.
\newblock


\bibitem[Fujita et~al\mbox{.}(2020)]%
        {196}
\bibfield{author}{\bibinfo{person}{Kazuyuki Fujita}, \bibinfo{person}{Daigo Hayashi}, \bibinfo{person}{Kotaro Hara}, \bibinfo{person}{Kazuki Takashima}, {and} \bibinfo{person}{Yoshifumi Kitamura}.} \bibinfo{year}{2020}\natexlab{}.
\newblock \showarticletitle{Techniques to Visualize Occluded Graph Elements for 2.5D Map Editing}. In \bibinfo{booktitle}{\emph{Extended Abstracts of the 2020 CHI Conference on Human Factors in Computing Systems}} (Honolulu, HI, USA) \emph{(\bibinfo{series}{CHI EA '20})}. \bibinfo{publisher}{Association for Computing Machinery}, \bibinfo{address}{New York, NY, USA}, \bibinfo{pages}{1–9}.
\newblock
\showISBNx{9781450368193}
\urldef\tempurl%
\url{https://doi.org/10.1145/3334480.3382987}
\showDOI{\tempurl}


\bibitem[FUKUMOTO(2009)]%
        {225}
\bibfield{author}{\bibinfo{person}{Masaaki FUKUMOTO}.} \bibinfo{year}{2009}\natexlab{}.
\newblock \showarticletitle{PuyoSheet and PuyoDots: simple techniques for adding "button-push" feeling to touch panels}. In \bibinfo{booktitle}{\emph{CHI '09 Extended Abstracts on Human Factors in Computing Systems}} (Boston, MA, USA) \emph{(\bibinfo{series}{CHI EA '09})}. \bibinfo{publisher}{Association for Computing Machinery}, \bibinfo{address}{New York, NY, USA}, \bibinfo{pages}{3925–3930}.
\newblock
\showISBNx{9781605582474}
\urldef\tempurl%
\url{https://doi.org/10.1145/1520340.1520595}
\showDOI{\tempurl}


\bibitem[Gui et~al\mbox{.}(2022)]%
        {98}
\bibfield{author}{\bibinfo{person}{Xinyue Gui}, \bibinfo{person}{Koki Toda}, \bibinfo{person}{Stela~Hanbyeol Seo}, \bibinfo{person}{Chia-Ming Chang}, {and} \bibinfo{person}{Takeo Igarashi}.} \bibinfo{year}{2022}\natexlab{}.
\newblock \showarticletitle{“I am going this way”: Gazing Eyes on Self-Driving Car Show Multiple Driving Directions}. In \bibinfo{booktitle}{\emph{Proceedings of the 14th International Conference on Automotive User Interfaces and Interactive Vehicular Applications}} (Seoul, Republic of Korea) \emph{(\bibinfo{series}{AutomotiveUI '22})}. \bibinfo{publisher}{Association for Computing Machinery}, \bibinfo{address}{New York, NY, USA}, \bibinfo{pages}{319–329}.
\newblock
\showISBNx{9781450394154}
\urldef\tempurl%
\url{https://doi.org/10.1145/3543174.3545251}
\showDOI{\tempurl}


\bibitem[Gui et~al\mbox{.}(2023)]%
        {232}
\bibfield{author}{\bibinfo{person}{Xinyue Gui}, \bibinfo{person}{Koki Toda}, \bibinfo{person}{Stela~Hanbyeol Seo}, \bibinfo{person}{Felix~Martin Eckert}, \bibinfo{person}{Chia-Ming Chang}, \bibinfo{person}{Xiang~'Anthony Chen}, {and} \bibinfo{person}{Takeo Igarashi}.} \bibinfo{year}{2023}\natexlab{}.
\newblock \showarticletitle{A Field Study on Pedestrians’ Thoughts toward a Car with Gazing Eyes}. In \bibinfo{booktitle}{\emph{Extended Abstracts of the 2023 CHI Conference on Human Factors in Computing Systems}} (Hamburg, Germany) \emph{(\bibinfo{series}{CHI EA '23})}. \bibinfo{publisher}{Association for Computing Machinery}, \bibinfo{address}{New York, NY, USA}, Article \bibinfo{articleno}{8}, \bibinfo{numpages}{7}~pages.
\newblock
\showISBNx{9781450394222}
\urldef\tempurl%
\url{https://doi.org/10.1145/3544549.3585629}
\showDOI{\tempurl}


\bibitem[Hashiguchi and Ogawa(2011)]%
        {174}
\bibfield{author}{\bibinfo{person}{Kyoko Hashiguchi} {and} \bibinfo{person}{Katsuhiko Ogawa}.} \bibinfo{year}{2011}\natexlab{}.
\newblock \showarticletitle{Proposal of the Kawaii Search System Based on the First Sight of Impression}. In \bibinfo{booktitle}{\emph{Human Interface and the Management of Information. Interacting with Information}}, \bibfield{editor}{\bibinfo{person}{Gavriel Salvendy} {and} \bibinfo{person}{Michael~J. Smith}} (Eds.). \bibinfo{publisher}{Springer Berlin Heidelberg}, \bibinfo{address}{Berlin, Heidelberg}, \bibinfo{pages}{21--30}.
\newblock
\showISBNx{978-3-642-21669-5}


\bibitem[Hatada et~al\mbox{.}(2019)]%
        {6}
\bibfield{author}{\bibinfo{person}{Yuji Hatada}, \bibinfo{person}{Shigeo Yoshida}, \bibinfo{person}{Takuji Narumi}, {and} \bibinfo{person}{Michitaka Hirose}.} \bibinfo{year}{2019}\natexlab{}.
\newblock \showarticletitle{Double Shellf: What Psychological Effects can be Caused through Interaction with a Doppelganger?}. In \bibinfo{booktitle}{\emph{Proceedings of the 10th Augmented Human International Conference 2019}} (Reims, France) \emph{(\bibinfo{series}{AH2019})}. \bibinfo{publisher}{Association for Computing Machinery}, \bibinfo{address}{New York, NY, USA}, Article \bibinfo{articleno}{34}, \bibinfo{numpages}{8}~pages.
\newblock
\showISBNx{9781450365475}
\urldef\tempurl%
\url{https://doi.org/10.1145/3311823.3311862}
\showDOI{\tempurl}


\bibitem[Hohendanner et~al\mbox{.}(2023)]%
        {39}
\bibfield{author}{\bibinfo{person}{Michel Hohendanner}, \bibinfo{person}{Chiara Ullstein}, \bibinfo{person}{Yosuke Buchmeier}, {and} \bibinfo{person}{Jens Grossklags}.} \bibinfo{year}{2023}\natexlab{}.
\newblock \showarticletitle{Exploring the Reflective Space of AI Narratives Through Speculative Design in Japan and Germany}. In \bibinfo{booktitle}{\emph{Proceedings of the 2023 ACM Conference on Information Technology for Social Good}} (Lisbon, Portugal) \emph{(\bibinfo{series}{GoodIT '23})}. \bibinfo{publisher}{Association for Computing Machinery}, \bibinfo{address}{New York, NY, USA}, \bibinfo{pages}{351–362}.
\newblock
\showISBNx{9798400701160}
\urldef\tempurl%
\url{https://doi.org/10.1145/3582515.3609554}
\showDOI{\tempurl}


\bibitem[Iwamoto et~al\mbox{.}(2022)]%
        {83}
\bibfield{author}{\bibinfo{person}{Takuya Iwamoto}, \bibinfo{person}{Jun Baba}, \bibinfo{person}{Junya Nakanishi}, \bibinfo{person}{Kotaro Nishi}, \bibinfo{person}{Yuichiro Yoshikawa}, {and} \bibinfo{person}{Hiroshi Ishiguro}.} \bibinfo{year}{2022}\natexlab{}.
\newblock \showarticletitle{Pick-me-up Strategy for a Self-recommendation Agent: A Pilot Field Experiment in a Convenience Store}. In \bibinfo{booktitle}{\emph{Proceedings of the 2022 ACM/IEEE International Conference on Human-Robot Interaction}} (Sapporo, Hokkaido, Japan) \emph{(\bibinfo{series}{HRI '22})}. \bibinfo{publisher}{IEEE}, \bibinfo{address}{New York, NY, USA}, \bibinfo{pages}{816–820}.
\newblock


\bibitem[Iwamoto et~al\mbox{.}(2021)]%
        {92}
\bibfield{author}{\bibinfo{person}{Takuya Iwamoto}, \bibinfo{person}{Jun Baba}, \bibinfo{person}{Kotaro Nishi}, \bibinfo{person}{Taishi Unokuchi}, \bibinfo{person}{Daisuke Endo}, \bibinfo{person}{Junya Nakanishi}, \bibinfo{person}{Yuichiro Yoshikawa}, {and} \bibinfo{person}{Hiroshi Ishiguro}.} \bibinfo{year}{2021}\natexlab{}.
\newblock \showarticletitle{The Effectiveness of Self-Recommending Agents in Advancing Purchase Behavior Steps in Retail Marketing}. In \bibinfo{booktitle}{\emph{Proceedings of the 9th International Conference on Human-Agent Interaction}} (Virtual Event, Japan) \emph{(\bibinfo{series}{HAI '21})}. \bibinfo{publisher}{Association for Computing Machinery}, \bibinfo{address}{New York, NY, USA}, \bibinfo{pages}{209–217}.
\newblock
\showISBNx{9781450386203}
\urldef\tempurl%
\url{https://doi.org/10.1145/3472307.3484183}
\showDOI{\tempurl}


\bibitem[Kanda et~al\mbox{.}(2023)]%
        {58}
\bibfield{author}{\bibinfo{person}{Shogo Kanda}, \bibinfo{person}{Masayuki Kanbara}, \bibinfo{person}{Taishi Sawabe}, \bibinfo{person}{Yuichiro Fujimoto}, {and} \bibinfo{person}{Hirokazu Kato}.} \bibinfo{year}{2023}\natexlab{}.
\newblock \showarticletitle{Robot to Play Video Games Together}. In \bibinfo{booktitle}{\emph{Proceedings of the 11th International Conference on Human-Agent Interaction}} (Gothenburg, Sweden) \emph{(\bibinfo{series}{HAI '23})}. \bibinfo{publisher}{Association for Computing Machinery}, \bibinfo{address}{New York, NY, USA}, \bibinfo{pages}{238–245}.
\newblock
\showISBNx{9798400708244}
\urldef\tempurl%
\url{https://doi.org/10.1145/3623809.3623832}
\showDOI{\tempurl}


\bibitem[Kanda et~al\mbox{.}(2009)]%
        {10}
\bibfield{author}{\bibinfo{person}{Takayuki Kanda}, \bibinfo{person}{Masahiro Shiomi}, \bibinfo{person}{Zenta Miyashita}, \bibinfo{person}{Hiroshi Ishiguro}, {and} \bibinfo{person}{Norihiro Hagita}.} \bibinfo{year}{2009}\natexlab{}.
\newblock \showarticletitle{An affective guide robot in a shopping mall}. In \bibinfo{booktitle}{\emph{Proceedings of the 4th ACM/IEEE International Conference on Human Robot Interaction}} (La Jolla, California, USA) \emph{(\bibinfo{series}{HRI '09})}. \bibinfo{publisher}{Association for Computing Machinery}, \bibinfo{address}{New York, NY, USA}, \bibinfo{pages}{173–180}.
\newblock
\showISBNx{9781605584041}
\urldef\tempurl%
\url{https://doi.org/10.1145/1514095.1514127}
\showDOI{\tempurl}


\bibitem[Kasuga et~al\mbox{.}(2017)]%
        {120}
\bibfield{author}{\bibinfo{person}{Haruka Kasuga}, \bibinfo{person}{Daisuke Sakamoto}, \bibinfo{person}{Nagisa Munekata}, {and} \bibinfo{person}{Tetsuo Ono}.} \bibinfo{year}{2017}\natexlab{}.
\newblock \showarticletitle{A Social Robot in a Human-Animal Relationship at Home: A Field Study}. In \bibinfo{booktitle}{\emph{Proceedings of the 5th International Conference on Human Agent Interaction}} (Bielefeld, Germany) \emph{(\bibinfo{series}{HAI '17})}. \bibinfo{publisher}{Association for Computing Machinery}, \bibinfo{address}{New York, NY, USA}, \bibinfo{pages}{61–69}.
\newblock
\showISBNx{9781450351133}
\urldef\tempurl%
\url{https://doi.org/10.1145/3125739.3125759}
\showDOI{\tempurl}


\bibitem[Kimura and Nakajima(2022)]%
        {211}
\bibfield{author}{\bibinfo{person}{Risa Kimura} {and} \bibinfo{person}{Tatsuo Nakajima}.} \bibinfo{year}{2022}\natexlab{}.
\newblock \showarticletitle{Hunting Lovely and Serendipitous Eye Sights through Sharing Collective Human Eye Views}. In \bibinfo{booktitle}{\emph{The 23rd International Conference on Information Integration and Web Intelligence}} (Linz, Austria) \emph{(\bibinfo{series}{iiWAS2021})}. \bibinfo{publisher}{Association for Computing Machinery}, \bibinfo{address}{New York, NY, USA}, \bibinfo{pages}{74–79}.
\newblock
\showISBNx{9781450395564}
\urldef\tempurl%
\url{https://doi.org/10.1145/3487664.3487675}
\showDOI{\tempurl}


\bibitem[Kinsella(2013)]%
        {kinsella2013cuties}
\bibfield{author}{\bibinfo{person}{Sharon Kinsella}.} \bibinfo{year}{2013}\natexlab{}.
\newblock \showarticletitle{Cuties in japan}.
\newblock In \bibinfo{booktitle}{\emph{Women, media and consumption in Japan}}. \bibinfo{publisher}{Routledge}, \bibinfo{address}{Milton Park Abingdon, UK}, \bibinfo{pages}{230--264}.
\newblock


\bibitem[Kobayashi et~al\mbox{.}(2019)]%
        {210}
\bibfield{author}{\bibinfo{person}{Kei Kobayashi}, \bibinfo{person}{Kazuma Nagata}, \bibinfo{person}{Soh Masuko}, {and} \bibinfo{person}{Junichi Hoshino}.} \bibinfo{year}{2019}\natexlab{}.
\newblock \showarticletitle{FUROSHIKI: Augmented Reality Media That Conveys Japanese Traditional Culture}. In \bibinfo{booktitle}{\emph{Proceedings of the 17th International Conference on Virtual-Reality Continuum and Its Applications in Industry}} (Brisbane, QLD, Australia) \emph{(\bibinfo{series}{VRCAI '19})}. \bibinfo{publisher}{Association for Computing Machinery}, \bibinfo{address}{New York, NY, USA}, Article \bibinfo{articleno}{27}, \bibinfo{numpages}{5}~pages.
\newblock
\showISBNx{9781450370028}
\urldef\tempurl%
\url{https://doi.org/10.1145/3359997.3365716}
\showDOI{\tempurl}


\bibitem[Koike and Itoh(2015)]%
        {122}
\bibfield{author}{\bibinfo{person}{Eriko Koike} {and} \bibinfo{person}{Takayuki Itoh}.} \bibinfo{year}{2015}\natexlab{}.
\newblock \showarticletitle{An interactive exploratory search system for on-line apparel shopping}. In \bibinfo{booktitle}{\emph{Proceedings of the 8th International Symposium on Visual Information Communication and Interaction}} (Tokyo, AA, Japan) \emph{(\bibinfo{series}{VINCI '15})}. \bibinfo{publisher}{Association for Computing Machinery}, \bibinfo{address}{New York, NY, USA}, \bibinfo{pages}{103–108}.
\newblock
\showISBNx{9781450334822}
\urldef\tempurl%
\url{https://doi.org/10.1145/2801040.2801041}
\showDOI{\tempurl}


\bibitem[Kovarovic(2011)]%
        {kovarovic2011hello}
\bibfield{author}{\bibinfo{person}{Sara Kovarovic}.} \bibinfo{year}{2011}\natexlab{}.
\newblock \showarticletitle{Hello Kitty: A brand made of cuteness}.
\newblock \bibinfo{journal}{\emph{Journal of Culture and Retail Image}} \bibinfo{volume}{4}, \bibinfo{number}{1} (\bibinfo{year}{2011}), \bibinfo{pages}{1--8}.
\newblock


\bibitem[Laohakangvalvit et~al\mbox{.}(2016)]%
        {159}
\bibfield{author}{\bibinfo{person}{Tipporn Laohakangvalvit}, \bibinfo{person}{Tiranee Achalakul}, {and} \bibinfo{person}{Michiko Ohkura}.} \bibinfo{year}{2016}\natexlab{}.
\newblock \showarticletitle{Kawaii feeling estimation by product attributes and biological signals}. In \bibinfo{booktitle}{\emph{Proceedings of the 18th ACM International Conference on Multimodal Interaction}} (Tokyo, Japan) \emph{(\bibinfo{series}{ICMI '16})}. \bibinfo{publisher}{Association for Computing Machinery}, \bibinfo{address}{New York, NY, USA}, \bibinfo{pages}{563–566}.
\newblock
\showISBNx{9781450345569}
\urldef\tempurl%
\url{https://doi.org/10.1145/2993148.2997621}
\showDOI{\tempurl}


\bibitem[Laohakangvalvit et~al\mbox{.}(2019)]%
        {148}
\bibfield{author}{\bibinfo{person}{Tipporn Laohakangvalvit}, \bibinfo{person}{Tiranee Achalakul}, {and} \bibinfo{person}{Michiko Ohkura}.} \bibinfo{year}{2019}\natexlab{}.
\newblock \showarticletitle{Comparison on Evaluation of Kawaiiness of Cosmetic Bottles between Japanese and Thai People}. In \bibinfo{booktitle}{\emph{2019 8th International Conference on Affective Computing and Intelligent Interaction (ACII)}}. \bibinfo{publisher}{IEEE}, \bibinfo{address}{New York, NY, USA}, \bibinfo{pages}{614--619}.
\newblock
\urldef\tempurl%
\url{https://doi.org/10.1109/ACII.2019.8925477}
\showDOI{\tempurl}


\bibitem[Laohakangvalvit et~al\mbox{.}(2022)]%
        {157}
\bibfield{author}{\bibinfo{person}{Tipporn Laohakangvalvit}, \bibinfo{person}{Peeraya Sripian}, \bibinfo{person}{Keiko Miyatake}, {and} \bibinfo{person}{Michiko Ohkura}.} \bibinfo{year}{2022}\natexlab{}.
\newblock \showarticletitle{A Proposal of Classification Model for Kawaii Fashion Styles in Japan Using Deep Learning}. In \bibinfo{booktitle}{\emph{Human-Computer Interaction. Theoretical Approaches and Design Methods}}, \bibfield{editor}{\bibinfo{person}{Masaaki Kurosu}} (Ed.). \bibinfo{publisher}{Springer International Publishing}, \bibinfo{address}{Cham}, \bibinfo{pages}{450--461}.
\newblock
\showISBNx{978-3-031-05311-5}


\bibitem[Laohakangvalvit et~al\mbox{.}(2021)]%
        {160}
\bibfield{author}{\bibinfo{person}{Tipporn Laohakangvalvit}, \bibinfo{person}{Peeraya Sripian}, \bibinfo{person}{Midori Sugaya}, {and} \bibinfo{person}{Michiko Ohkura}.} \bibinfo{year}{2021}\natexlab{}.
\newblock \showarticletitle{Relationship Between Robot Designs and Preferences in Kawaii Attributes}. In \bibinfo{booktitle}{\emph{Human-Computer Interaction. Interaction Techniques and Novel Applications}}, \bibfield{editor}{\bibinfo{person}{Masaaki Kurosu}} (Ed.). \bibinfo{publisher}{Springer International Publishing}, \bibinfo{address}{Cham}, \bibinfo{pages}{251--261}.
\newblock
\showISBNx{978-3-030-78465-2}


\bibitem[Lieber-Milo and Nittono(2019)]%
        {lieber2019word}
\bibfield{author}{\bibinfo{person}{Shiri Lieber-Milo} {and} \bibinfo{person}{Hiroshi Nittono}.} \bibinfo{year}{2019}\natexlab{}.
\newblock \showarticletitle{From a word to a commercial power: A brief introduction to the kawaii aesthetic in contemporary Japan}.
\newblock \bibinfo{journal}{\emph{Innovative Research in Japanese Studies}}  \bibinfo{volume}{3} (\bibinfo{year}{2019}), \bibinfo{pages}{13--32}.
\newblock


\bibitem[Lorenz(1943)]%
        {lorenz1943innate}
\bibfield{author}{\bibinfo{person}{Konrad Lorenz}.} \bibinfo{year}{1943}\natexlab{}.
\newblock \showarticletitle{The innate forms of potential experience}.
\newblock \bibinfo{journal}{\emph{Z Tierpsychol}}  \bibinfo{volume}{5} (\bibinfo{year}{1943}), \bibinfo{pages}{235}.
\newblock


\bibitem[Lu et~al\mbox{.}(2021)]%
        {2}
\bibfield{author}{\bibinfo{person}{Zhicong Lu}, \bibinfo{person}{Chenxinran Shen}, \bibinfo{person}{Jiannan Li}, \bibinfo{person}{Hong Shen}, {and} \bibinfo{person}{Daniel Wigdor}.} \bibinfo{year}{2021}\natexlab{}.
\newblock \showarticletitle{More Kawaii than a Real-Person Live Streamer: Understanding How the Otaku Community Engages with and Perceives Virtual YouTubers}. In \bibinfo{booktitle}{\emph{Proceedings of the 2021 CHI Conference on Human Factors in Computing Systems}} (Yokohama, Japan) \emph{(\bibinfo{series}{CHI '21})}. \bibinfo{publisher}{Association for Computing Machinery}, \bibinfo{address}{New York, NY, USA}, Article \bibinfo{articleno}{137}, \bibinfo{numpages}{14}~pages.
\newblock
\showISBNx{9781450380966}
\urldef\tempurl%
\url{https://doi.org/10.1145/3411764.3445660}
\showDOI{\tempurl}


\bibitem[Marcus et~al\mbox{.}(2017)]%
        {marcus2017cuteness}
\bibfield{author}{\bibinfo{person}{Aaron Marcus}, \bibinfo{person}{Masaaki Kurosu}, \bibinfo{person}{Xiaojuan Ma}, {and} \bibinfo{person}{Ayako Hashizume}.} \bibinfo{year}{2017}\natexlab{}.
\newblock \bibinfo{booktitle}{\emph{Cuteness engineering: Designing adorable products and services}}.
\newblock \bibinfo{publisher}{Springer}, \bibinfo{address}{London, UK}.
\newblock


\bibitem[Mashimo et~al\mbox{.}(2016)]%
        {130}
\bibfield{author}{\bibinfo{person}{Ryo Mashimo}, \bibinfo{person}{Tomohiro Umetani}, \bibinfo{person}{Tatsuya Kitamura}, {and} \bibinfo{person}{Akiyo Nadamoto}.} \bibinfo{year}{2016}\natexlab{}.
\newblock \showarticletitle{Human-Robots Implicit Communication based on Dialogue between Robots using Automatic Generation of Funny Scenarios from Web}. In \bibinfo{booktitle}{\emph{The Eleventh ACM/IEEE International Conference on Human Robot Interaction}} (Christchurch, New Zealand) \emph{(\bibinfo{series}{HRI '16})}. \bibinfo{publisher}{IEEE}, \bibinfo{address}{New York, NY, USA}, \bibinfo{pages}{327–334}.
\newblock
\showISBNx{9781467383707}


\bibitem[Matsui and Yamada(2016)]%
        {18}
\bibfield{author}{\bibinfo{person}{Tetsuya Matsui} {and} \bibinfo{person}{Seiji Yamada}.} \bibinfo{year}{2016}\natexlab{}.
\newblock \showarticletitle{Building Trust in PRVAs by User Inner State Transition through Agent State Transition}. In \bibinfo{booktitle}{\emph{Proceedings of the Fourth International Conference on Human Agent Interaction}} (Biopolis, Singapore) \emph{(\bibinfo{series}{HAI '16})}. \bibinfo{publisher}{Association for Computing Machinery}, \bibinfo{address}{New York, NY, USA}, \bibinfo{pages}{111–114}.
\newblock
\showISBNx{9781450345088}
\urldef\tempurl%
\url{https://doi.org/10.1145/2974804.2974816}
\showDOI{\tempurl}


\bibitem[Minakawa and Takada(2017)]%
        {205}
\bibfield{author}{\bibinfo{person}{Ryo Minakawa} {and} \bibinfo{person}{Tetsuji Takada}.} \bibinfo{year}{2017}\natexlab{}.
\newblock \showarticletitle{Exploring alternative security warning dialog for attracting user attention: evaluation of "Kawaii" effect and its additional stimulus combination}. In \bibinfo{booktitle}{\emph{Proceedings of the 19th International Conference on Information Integration and Web-Based Applications \& Services}} (Salzburg, Austria) \emph{(\bibinfo{series}{iiWAS '17})}. \bibinfo{publisher}{Association for Computing Machinery}, \bibinfo{address}{New York, NY, USA}, \bibinfo{pages}{582–586}.
\newblock
\showISBNx{9781450352994}
\urldef\tempurl%
\url{https://doi.org/10.1145/3151759.3151846}
\showDOI{\tempurl}


\bibitem[Mizuho et~al\mbox{.}(2023)]%
        {71}
\bibfield{author}{\bibinfo{person}{Takato Mizuho}, \bibinfo{person}{Tomohiro Amemiya}, \bibinfo{person}{Takuji Narumi}, {and} \bibinfo{person}{Hideaki Kuzuoka}.} \bibinfo{year}{2023}\natexlab{}.
\newblock \showarticletitle{Virtual Omnibus Lecture: Investigating the Effects of Varying Lecturer Avatars as Environmental Context on Audience Memory}. In \bibinfo{booktitle}{\emph{Proceedings of the Augmented Humans International Conference 2023}} (Glasgow, United Kingdom) \emph{(\bibinfo{series}{AHs '23})}. \bibinfo{publisher}{Association for Computing Machinery}, \bibinfo{address}{New York, NY, USA}, \bibinfo{pages}{55–65}.
\newblock
\showISBNx{9781450399845}
\urldef\tempurl%
\url{https://doi.org/10.1145/3582700.3582709}
\showDOI{\tempurl}


\bibitem[Mubin et~al\mbox{.}(2020)]%
        {230}
\bibfield{author}{\bibinfo{person}{Omar Mubin}, \bibinfo{person}{Isha Kharub}, {and} \bibinfo{person}{Aila Khan}.} \bibinfo{year}{2020}\natexlab{}.
\newblock \showarticletitle{Pepper in the Library" Students' First Impressions}. In \bibinfo{booktitle}{\emph{Extended Abstracts of the 2020 CHI Conference on Human Factors in Computing Systems}} (Honolulu, HI, USA) \emph{(\bibinfo{series}{CHI EA '20})}. \bibinfo{publisher}{Association for Computing Machinery}, \bibinfo{address}{New York, NY, USA}, \bibinfo{pages}{1–9}.
\newblock
\showISBNx{9781450368193}
\urldef\tempurl%
\url{https://doi.org/10.1145/3334480.3382979}
\showDOI{\tempurl}


\bibitem[Munn et~al\mbox{.}(2018)]%
        {Munn2018}
\bibfield{author}{\bibinfo{person}{Zachary Munn}, \bibinfo{person}{Micah D.~J. Peters}, \bibinfo{person}{Cindy Stern}, \bibinfo{person}{Catalin Tufanaru}, \bibinfo{person}{Alexa McArthur}, {and} \bibinfo{person}{Edoardo Aromataris}.} \bibinfo{year}{2018}\natexlab{}.
\newblock \showarticletitle{Systematic review or scoping review? Guidance for authors when choosing between a systematic or scoping review approach}.
\newblock \bibinfo{journal}{\emph{BMC Medical Research Methodology}} \bibinfo{volume}{18}, \bibinfo{number}{1} (\bibinfo{year}{2018}), \bibinfo{numpages}{7}~pages.
\newblock
\showISSN{1471-2288}
\urldef\tempurl%
\url{https://doi.org/10.1186/s12874-018-0611-x}
\showDOI{\tempurl}


\bibitem[Nakagawa et~al\mbox{.}(2011)]%
        {57}
\bibfield{author}{\bibinfo{person}{Kayako Nakagawa}, \bibinfo{person}{Masahiro Shiomi}, \bibinfo{person}{Kazuhiko Shinozawa}, \bibinfo{person}{Reo Matsumura}, \bibinfo{person}{Hiroshi Ishiguro}, {and} \bibinfo{person}{Norihiro Hagita}.} \bibinfo{year}{2011}\natexlab{}.
\newblock \showarticletitle{Effect of robot's active touch on people's motivation}. In \bibinfo{booktitle}{\emph{Proceedings of the 6th International Conference on Human-Robot Interaction}} (Lausanne, Switzerland) \emph{(\bibinfo{series}{HRI '11})}. \bibinfo{publisher}{Association for Computing Machinery}, \bibinfo{address}{New York, NY, USA}, \bibinfo{pages}{465–472}.
\newblock
\showISBNx{9781450305617}
\urldef\tempurl%
\url{https://doi.org/10.1145/1957656.1957819}
\showDOI{\tempurl}


\bibitem[Nakajima and Niitsuma(2020)]%
        {8}
\bibfield{author}{\bibinfo{person}{Kenta Nakajima} {and} \bibinfo{person}{Mihoko Niitsuma}.} \bibinfo{year}{2020}\natexlab{}.
\newblock \showarticletitle{Effects of Space and Scenery on Virtual Pet-Assisted Activity}. In \bibinfo{booktitle}{\emph{Proceedings of the 8th International Conference on Human-Agent Interaction}} (Virtual Event, USA) \emph{(\bibinfo{series}{HAI '20})}. \bibinfo{publisher}{Association for Computing Machinery}, \bibinfo{address}{New York, NY, USA}, \bibinfo{pages}{105–111}.
\newblock
\showISBNx{9781450380546}
\urldef\tempurl%
\url{https://doi.org/10.1145/3406499.3415083}
\showDOI{\tempurl}


\bibitem[Nakanishi et~al\mbox{.}(2022)]%
        {52}
\bibfield{author}{\bibinfo{person}{Junya Nakanishi}, \bibinfo{person}{Jun Baba}, {and} \bibinfo{person}{Hiroshi Ishiguro}.} \bibinfo{year}{2022}\natexlab{}.
\newblock \showarticletitle{Robot-Mediated Interaction Between Children and Older Adults: A Pilot Study for Greeting Tasks in Nursery Schools}. In \bibinfo{booktitle}{\emph{Proceedings of the 2022 ACM/IEEE International Conference on Human-Robot Interaction}} (Sapporo, Hokkaido, Japan) \emph{(\bibinfo{series}{HRI '22})}. \bibinfo{publisher}{IEEE}, \bibinfo{address}{New York, NY, USA}, \bibinfo{pages}{63–70}.
\newblock


\bibitem[Nakanishi et~al\mbox{.}(2019)]%
        {218}
\bibfield{author}{\bibinfo{person}{Junya Nakanishi}, \bibinfo{person}{Jun Baba}, {and} \bibinfo{person}{Itaru Kuramoto}.} \bibinfo{year}{2019}\natexlab{}.
\newblock \showarticletitle{How to Enhance Social Robots' Heartwarming Interaction in Service Encounters}. In \bibinfo{booktitle}{\emph{Proceedings of the 7th International Conference on Human-Agent Interaction}} (Kyoto, Japan) \emph{(\bibinfo{series}{HAI '19})}. \bibinfo{publisher}{Association for Computing Machinery}, \bibinfo{address}{New York, NY, USA}, \bibinfo{pages}{297–299}.
\newblock
\showISBNx{9781450369220}
\urldef\tempurl%
\url{https://doi.org/10.1145/3349537.3352798}
\showDOI{\tempurl}


\bibitem[Nakanishi et~al\mbox{.}(2018)]%
        {134}
\bibfield{author}{\bibinfo{person}{Junya Nakanishi}, \bibinfo{person}{Itaru Kuramoto}, \bibinfo{person}{Jun Baba}, \bibinfo{person}{Ogawa Kohei}, \bibinfo{person}{Yuichiro Yoshikawa}, {and} \bibinfo{person}{Hiroshi Ishiguro}.} \bibinfo{year}{2018}\natexlab{}.
\newblock \showarticletitle{Can a Humanoid Robot Engage in Heartwarming Interaction Service at a Hotel?}. In \bibinfo{booktitle}{\emph{Proceedings of the 6th International Conference on Human-Agent Interaction}} (Southampton, United Kingdom) \emph{(\bibinfo{series}{HAI '18})}. \bibinfo{publisher}{Association for Computing Machinery}, \bibinfo{address}{New York, NY, USA}, \bibinfo{pages}{45–53}.
\newblock
\showISBNx{9781450359535}
\urldef\tempurl%
\url{https://doi.org/10.1145/3284432.3284448}
\showDOI{\tempurl}


\bibitem[Nature(2017)]%
        {natureopenscience2017}
\bibfield{author}{\bibinfo{person}{Nature}.} \bibinfo{year}{2017}\natexlab{}.
\newblock \showarticletitle{Open science}.
\newblock \bibinfo{journal}{\emph{Nature}} \bibinfo{volume}{550}, \bibinfo{number}{7674} (\bibinfo{date}{Oct.} \bibinfo{year}{2017}), \bibinfo{pages}{7–8}.
\newblock
\showISSN{1476-4687}
\urldef\tempurl%
\url{https://doi.org/10.1038/550007b}
\showDOI{\tempurl}


\bibitem[Nittono(2010)]%
        {nittono2010behavioral}
\bibfield{author}{\bibinfo{person}{Hiroshi Nittono}.} \bibinfo{year}{2010}\natexlab{}.
\newblock \showarticletitle{A behavioral science framework for understanding kawaii}. In \bibinfo{booktitle}{\emph{Proceedings of The Third International Workshop on Kansei}}. Editorial comittee of the third international workshop on Kansei, \bibinfo{publisher}{Japanese Society for Cognitive Psychology}, \bibinfo{address}{Nishi, Fukuoka, Japan}, \bibinfo{pages}{80--83}.
\newblock


\bibitem[Nittono(2016)]%
        {nittono2016two}
\bibfield{author}{\bibinfo{person}{Hiroshi Nittono}.} \bibinfo{year}{2016}\natexlab{}.
\newblock \showarticletitle{The two-layer model of ‘kawaii’: A behavioural science framework for understanding kawaii and cuteness}.
\newblock \bibinfo{journal}{\emph{East Asian Journal of Popular Culture}} \bibinfo{volume}{2}, \bibinfo{number}{1} (\bibinfo{year}{2016}), \bibinfo{pages}{79--95}.
\newblock


\bibitem[Nittono(2019)]%
        {nittono2019kawaii}
\bibfield{author}{\bibinfo{person}{Hiroshi Nittono}.} \bibinfo{year}{2019}\natexlab{}.
\newblock \bibinfo{booktitle}{\emph{Kawaii no Chikara (The Power of Kawaii)}}.
\newblock \bibinfo{publisher}{Dojin Sensho}, \bibinfo{address}{Kyoto, Japan}.
\newblock


\bibitem[Nittono(2022)]%
        {nittono2022psychology}
\bibfield{author}{\bibinfo{person}{Hiroshi Nittono}.} \bibinfo{year}{2022}\natexlab{}.
\newblock \showarticletitle{The Psychology of “Kawaii” and Its Implications for Human-Robot Interaction}. In \bibinfo{booktitle}{\emph{2022 17th ACM/IEEE International Conference on Human-Robot Interaction (HRI)}}. \bibinfo{publisher}{IEEE}, \bibinfo{address}{New York, NY, USA}, \bibinfo{pages}{3--3}.
\newblock


\bibitem[Nittono et~al\mbox{.}(2012)]%
        {Nittono2012}
\bibfield{author}{\bibinfo{person}{Hiroshi Nittono}, \bibinfo{person}{Michiko Fukushima}, \bibinfo{person}{Akihiro Yano}, {and} \bibinfo{person}{Hiroki Moriya}.} \bibinfo{year}{2012}\natexlab{}.
\newblock \showarticletitle{The Power of Kawaii: Viewing Cute Images Promotes a Careful Behavior and Narrows Attentional Focus}.
\newblock \bibinfo{journal}{\emph{PLoS ONE}} \bibinfo{volume}{7}, \bibinfo{number}{9} (\bibinfo{date}{Sept.} \bibinfo{year}{2012}), \bibinfo{pages}{e46362}.
\newblock
\showISSN{1932-6203}
\urldef\tempurl%
\url{https://doi.org/10.1371/journal.pone.0046362}
\showDOI{\tempurl}


\bibitem[Noguchi et~al\mbox{.}(2021)]%
        {215}
\bibfield{author}{\bibinfo{person}{Kureha Noguchi}, \bibinfo{person}{Yoshinari Takegawa}, \bibinfo{person}{Yutaka Tokuda}, \bibinfo{person}{Yuta Sugiura}, \bibinfo{person}{Katsutoshi Masai}, {and} \bibinfo{person}{Keiji Hirata}.} \bibinfo{year}{2021}\natexlab{}.
\newblock \showarticletitle{Study of Interviewee’s ImpressionMade by Interviewer Wearing Digital Full-face Mask DisplayDuring Recruitment Interview}. In \bibinfo{booktitle}{\emph{Proceedings of the 9th International Conference on Human-Agent Interaction}} (Virtual Event, Japan) \emph{(\bibinfo{series}{HAI '21})}. \bibinfo{publisher}{Association for Computing Machinery}, \bibinfo{address}{New York, NY, USA}, \bibinfo{pages}{323–327}.
\newblock
\showISBNx{9781450386203}
\urldef\tempurl%
\url{https://doi.org/10.1145/3472307.3484662}
\showDOI{\tempurl}


\bibitem[Ohkubo et~al\mbox{.}(2016)]%
        {208}
\bibfield{author}{\bibinfo{person}{Masaru Ohkubo}, \bibinfo{person}{Shuhei Umezu}, {and} \bibinfo{person}{Takuya Nojima}.} \bibinfo{year}{2016}\natexlab{}.
\newblock \showarticletitle{Come alive! Augmented Mobile Interaction with Smart Hair}. In \bibinfo{booktitle}{\emph{Proceedings of the 7th Augmented Human International Conference 2016}} (Geneva, Switzerland) \emph{(\bibinfo{series}{AH '16})}. \bibinfo{publisher}{Association for Computing Machinery}, \bibinfo{address}{New York, NY, USA}, Article \bibinfo{articleno}{32}, \bibinfo{numpages}{4}~pages.
\newblock
\showISBNx{9781450336802}
\urldef\tempurl%
\url{https://doi.org/10.1145/2875194.2875241}
\showDOI{\tempurl}


\bibitem[Ohkura(2019)]%
        {ohkura2019kawaii}
\bibfield{author}{\bibinfo{person}{Michiko Ohkura}.} \bibinfo{year}{2019}\natexlab{}.
\newblock \bibinfo{booktitle}{\emph{Kawaii engineering: Measurements, evaluations, and applications of attractiveness}}.
\newblock \bibinfo{publisher}{Springer}, \bibinfo{address}{London, UK}.
\newblock


\bibitem[Ohkura and Aoto(2010)]%
        {ohkura2010systematic}
\bibfield{author}{\bibinfo{person}{Michiko Ohkura} {and} \bibinfo{person}{Tetsuro Aoto}.} \bibinfo{year}{2010}\natexlab{}.
\newblock \showarticletitle{Systematic study of kawaii products: Relation between kawaii feelings and attributes of industrial products}. In \bibinfo{booktitle}{\emph{International Design Engineering Technical Conferences and Computers and Information in Engineering Conference}}, Vol.~\bibinfo{volume}{44113}. \bibinfo{publisher}{ASME}, \bibinfo{address}{New York, NY, USA}, \bibinfo{pages}{587--594}.
\newblock


\bibitem[Ohkura et~al\mbox{.}(2014)]%
        {ohkura2014kawaii}
\bibfield{author}{\bibinfo{person}{Michiko Ohkura}, \bibinfo{person}{Tsuyoshi Komatsu}, {and} \bibinfo{person}{Tetsuro Aoto}.} \bibinfo{year}{2014}\natexlab{}.
\newblock \showarticletitle{Kawaii rules: increasing affective value of industrial products}. In \bibinfo{booktitle}{\emph{Industrial applications of affective engineering}}. \bibinfo{publisher}{Springer}, \bibinfo{address}{London, UK}, \bibinfo{pages}{97--110}.
\newblock


\bibitem[Ohkura et~al\mbox{.}(2022)]%
        {161}
\bibfield{author}{\bibinfo{person}{Michiko Ohkura}, \bibinfo{person}{Tipporn Laohakangvalvit}, \bibinfo{person}{Peeraya Sripian}, \bibinfo{person}{Midori Sugaya}, \bibinfo{person}{Hiroko Chiba}, {and} \bibinfo{person}{Dave Berque}.} \bibinfo{year}{2022}\natexlab{}.
\newblock \showarticletitle{Affective Evaluation of Virtual Kawaii Robotic Gadgets Using Biological Signals in a Remote Collaboration of American and Japanese Students}. In \bibinfo{booktitle}{\emph{Human-Computer Interaction. Theoretical Approaches and Design Methods}}, \bibfield{editor}{\bibinfo{person}{Masaaki Kurosu}} (Ed.). \bibinfo{publisher}{Springer International Publishing}, \bibinfo{address}{Cham}, \bibinfo{pages}{478--488}.
\newblock
\showISBNx{978-3-031-05311-5}


\bibitem[Ohtsuka et~al\mbox{.}(2022)]%
        {162}
\bibfield{author}{\bibinfo{person}{Shuma Ohtsuka}, \bibinfo{person}{Peeraya Sripian}, \bibinfo{person}{Tipporn Laohakangvalvit}, {and} \bibinfo{person}{Midori Sugaya}.} \bibinfo{year}{2022}\natexlab{}.
\newblock \showarticletitle{Model Construction of ``Kawaii Characters'' Using Deep Learning}. In \bibinfo{booktitle}{\emph{Human-Computer Interaction. Theoretical Approaches and Design Methods}}, \bibfield{editor}{\bibinfo{person}{Masaaki Kurosu}} (Ed.). \bibinfo{publisher}{Springer International Publishing}, \bibinfo{address}{Cham}, \bibinfo{pages}{502--510}.
\newblock
\showISBNx{978-3-031-05311-5}


\bibitem[Okada et~al\mbox{.}(2020)]%
        {149}
\bibfield{author}{\bibinfo{person}{Yuka Okada}, \bibinfo{person}{Mitsuhiko Kimoto}, \bibinfo{person}{Takamasa Iio}, \bibinfo{person}{Katsunori Shimohara}, \bibinfo{person}{Hiroshi Nittono}, {and} \bibinfo{person}{Masahiro Shiomi}.} \bibinfo{year}{2020}\natexlab{}.
\newblock \showarticletitle{Can a Robot's Touches Express the Feeling of Kawaii toward an Object?}. In \bibinfo{booktitle}{\emph{2020 IEEE/RSJ International Conference on Intelligent Robots and Systems (IROS)}}. \bibinfo{publisher}{IEEE}, \bibinfo{address}{New York, NY, USA}, \bibinfo{pages}{11276--11283}.
\newblock
\urldef\tempurl%
\url{https://doi.org/10.1109/IROS45743.2020.9340874}
\showDOI{\tempurl}


\bibitem[Ootsubo and Inoue(2022)]%
        {213}
\bibfield{author}{\bibinfo{person}{Kaito Ootsubo} {and} \bibinfo{person}{Akifumi Inoue}.} \bibinfo{year}{2022}\natexlab{}.
\newblock \showarticletitle{Early Implementation of VR Stuffed Toy System with Virtual Softness}. In \bibinfo{booktitle}{\emph{Proceedings of the 33rd Australian Conference on Human-Computer Interaction}} (Melbourne, VIC, Australia) \emph{(\bibinfo{series}{OzCHI '21})}. \bibinfo{publisher}{Association for Computing Machinery}, \bibinfo{address}{New York, NY, USA}, \bibinfo{pages}{265–268}.
\newblock
\showISBNx{9781450395984}
\urldef\tempurl%
\url{https://doi.org/10.1145/3520495.3520499}
\showDOI{\tempurl}


\bibitem[Patrick~Rau et~al\mbox{.}(2018)]%
        {172}
\bibfield{author}{\bibinfo{person}{Pei-Luen Patrick~Rau}, \bibinfo{person}{Nan Qie}, {and} \bibinfo{person}{Chien-Wen Tung}.} \bibinfo{year}{2018}\natexlab{}.
\newblock \showarticletitle{Are Kawaii Products Valuable to Chinese Customers?}. In \bibinfo{booktitle}{\emph{Advances in Affective and Pleasurable Design}}, \bibfield{editor}{\bibinfo{person}{WonJoon Chung} {and} \bibinfo{person}{Cliff~Sungsoo Shin}} (Eds.). \bibinfo{publisher}{Springer International Publishing}, \bibinfo{address}{Cham}, \bibinfo{pages}{258--265}.
\newblock
\showISBNx{978-3-319-60495-4}


\bibitem[Poeller et~al\mbox{.}(2021)]%
        {63}
\bibfield{author}{\bibinfo{person}{Susanne Poeller}, \bibinfo{person}{Karla Waldenmeier}, \bibinfo{person}{Nicola Baumann}, {and} \bibinfo{person}{Regan Mandryk}.} \bibinfo{year}{2021}\natexlab{}.
\newblock \showarticletitle{Prepare for Trouble and Make It Double: The Power Motive Predicts Pok\'{e}mon Choices Based on Apparent Strength}. In \bibinfo{booktitle}{\emph{Proceedings of the 2021 CHI Conference on Human Factors in Computing Systems}} (Yokohama, Japan) \emph{(\bibinfo{series}{CHI '21})}. \bibinfo{publisher}{Association for Computing Machinery}, \bibinfo{address}{New York, NY, USA}, Article \bibinfo{articleno}{110}, \bibinfo{numpages}{12}~pages.
\newblock
\showISBNx{9781450380966}
\urldef\tempurl%
\url{https://doi.org/10.1145/3411764.3445084}
\showDOI{\tempurl}


\bibitem[Riessland(1998)]%
        {riessland1998sweet}
\bibfield{author}{\bibinfo{person}{Andreas Riessland}.} \bibinfo{year}{1998}\natexlab{}.
\newblock \showarticletitle{Sweet spots: The use of cuteness in Japanese advertising}.
\newblock \bibinfo{journal}{\emph{Japanstudien}} \bibinfo{volume}{9}, \bibinfo{number}{1} (\bibinfo{year}{1998}), \bibinfo{pages}{129--154}.
\newblock


\bibitem[Sabelli et~al\mbox{.}(2011)]%
        {65}
\bibfield{author}{\bibinfo{person}{Alessandra~Maria Sabelli}, \bibinfo{person}{Takayuki Kanda}, {and} \bibinfo{person}{Norihiro Hagita}.} \bibinfo{year}{2011}\natexlab{}.
\newblock \showarticletitle{A conversational robot in an elderly care center: an ethnographic study}. In \bibinfo{booktitle}{\emph{Proceedings of the 6th International Conference on Human-Robot Interaction}} (Lausanne, Switzerland) \emph{(\bibinfo{series}{HRI '11})}. \bibinfo{publisher}{Association for Computing Machinery}, \bibinfo{address}{New York, NY, USA}, \bibinfo{pages}{37–44}.
\newblock
\showISBNx{9781450305617}
\urldef\tempurl%
\url{https://doi.org/10.1145/1957656.1957669}
\showDOI{\tempurl}


\bibitem[Saga et~al\mbox{.}(2014)]%
        {91}
\bibfield{author}{\bibinfo{person}{Tamami Saga}, \bibinfo{person}{Nagisa Munekata}, {and} \bibinfo{person}{Tetsuo Ono}.} \bibinfo{year}{2014}\natexlab{}.
\newblock \showarticletitle{Daily support robots that move on the body}. In \bibinfo{booktitle}{\emph{Proceedings of the Second International Conference on Human-Agent Interaction}} (Tsukuba, Japan) \emph{(\bibinfo{series}{HAI '14})}. \bibinfo{publisher}{Association for Computing Machinery}, \bibinfo{address}{New York, NY, USA}, \bibinfo{pages}{29–34}.
\newblock
\showISBNx{9781450330350}
\urldef\tempurl%
\url{https://doi.org/10.1145/2658861.2658882}
\showDOI{\tempurl}


\bibitem[Sajadieh(2023)]%
        {176}
\bibfield{author}{\bibinfo{person}{Sahar Sajadieh}.} \bibinfo{year}{2023}\natexlab{}.
\newblock \showarticletitle{Cute or creepy, that is the question of liveness: can artificial actors perform live?}
\newblock \bibinfo{journal}{\emph{Artnodes}} \bibinfo{number}{32} (\bibinfo{date}{Jul.} \bibinfo{year}{2023}), \bibinfo{pages}{1--9}.
\newblock
\urldef\tempurl%
\url{https://doi.org/10.7238/artnodes.v0i32.412093}
\showDOI{\tempurl}


\bibitem[Sakamoto et~al\mbox{.}(2023)]%
        {80}
\bibfield{author}{\bibinfo{person}{Yumiko Sakamoto}, \bibinfo{person}{Anuradha Herath}, \bibinfo{person}{Tanvi Vuradi}, \bibinfo{person}{Samar Sallam}, \bibinfo{person}{Randy Gomez}, {and} \bibinfo{person}{Pourang Irani}.} \bibinfo{year}{2023}\natexlab{}.
\newblock \showarticletitle{How Should a Social Robot Deliver Negative Feedback Without Creating Distance Between the Robot and Child Users?}. In \bibinfo{booktitle}{\emph{Proceedings of the 11th International Conference on Human-Agent Interaction}} (Gothenburg, Sweden) \emph{(\bibinfo{series}{HAI '23})}. \bibinfo{publisher}{Association for Computing Machinery}, \bibinfo{address}{New York, NY, USA}, \bibinfo{pages}{325–334}.
\newblock
\showISBNx{9798400708244}
\urldef\tempurl%
\url{https://doi.org/10.1145/3623809.3623882}
\showDOI{\tempurl}


\bibitem[Sakashita et~al\mbox{.}(2017)]%
        {81}
\bibfield{author}{\bibinfo{person}{Mose Sakashita}, \bibinfo{person}{Tatsuya Minagawa}, \bibinfo{person}{Amy Koike}, \bibinfo{person}{Ippei Suzuki}, \bibinfo{person}{Keisuke Kawahara}, {and} \bibinfo{person}{Yoichi Ochiai}.} \bibinfo{year}{2017}\natexlab{}.
\newblock \showarticletitle{You as a Puppet: Evaluation of Telepresence User Interface for Puppetry}. In \bibinfo{booktitle}{\emph{Proceedings of the 30th Annual ACM Symposium on User Interface Software and Technology}} (Qu\'{e}bec City, QC, Canada) \emph{(\bibinfo{series}{UIST '17})}. \bibinfo{publisher}{Association for Computing Machinery}, \bibinfo{address}{New York, NY, USA}, \bibinfo{pages}{217–228}.
\newblock
\showISBNx{9781450349819}
\urldef\tempurl%
\url{https://doi.org/10.1145/3126594.3126608}
\showDOI{\tempurl}


\bibitem[Samani et~al\mbox{.}(2012)]%
        {90}
\bibfield{author}{\bibinfo{person}{Hooman~Aghaebrahimi Samani}, \bibinfo{person}{Rahul Parsani}, \bibinfo{person}{Lenis~Tejada Rodriguez}, \bibinfo{person}{Elham Saadatian}, \bibinfo{person}{Kumudu~Harshadeva Dissanayake}, {and} \bibinfo{person}{Adrian~David Cheok}.} \bibinfo{year}{2012}\natexlab{}.
\newblock \showarticletitle{Kissenger: design of a kiss transmission device}. In \bibinfo{booktitle}{\emph{Proceedings of the Designing Interactive Systems Conference}} (Newcastle Upon Tyne, United Kingdom) \emph{(\bibinfo{series}{DIS '12})}. \bibinfo{publisher}{Association for Computing Machinery}, \bibinfo{address}{New York, NY, USA}, \bibinfo{pages}{48–57}.
\newblock
\showISBNx{9781450312103}
\urldef\tempurl%
\url{https://doi.org/10.1145/2317956.2317965}
\showDOI{\tempurl}


\bibitem[Scissors et~al\mbox{.}(2011)]%
        {142}
\bibfield{author}{\bibinfo{person}{Lauren Scissors}, \bibinfo{person}{N.~Sadat Shami}, \bibinfo{person}{Tatsuya Ishihara}, \bibinfo{person}{Steven Rohall}, {and} \bibinfo{person}{Shin Saito}.} \bibinfo{year}{2011}\natexlab{}.
\newblock \showarticletitle{Real-time collaborative editing behavior in USA and Japanese distributed teams}. In \bibinfo{booktitle}{\emph{Proceedings of the SIGCHI Conference on Human Factors in Computing Systems}} (Vancouver, BC, Canada) \emph{(\bibinfo{series}{CHI '11})}. \bibinfo{publisher}{Association for Computing Machinery}, \bibinfo{address}{New York, NY, USA}, \bibinfo{pages}{1119–1128}.
\newblock
\showISBNx{9781450302289}
\urldef\tempurl%
\url{https://doi.org/10.1145/1978942.1979109}
\showDOI{\tempurl}


\bibitem[Seaborn et~al\mbox{.}(2023a)]%
        {seaborncanvoicesoundcute}
\bibfield{author}{\bibinfo{person}{Katie Seaborn}, \bibinfo{person}{Somang Nam}, \bibinfo{person}{Julia Keckeis}, {and} \bibinfo{person}{Tatsuya Itagaki}.} \bibinfo{year}{2023}\natexlab{a}.
\newblock \showarticletitle{Can Voice Assistants Sound Cute? Towards a Model of Kawaii Vocalics}. In \bibinfo{booktitle}{\emph{Extended Abstracts of the 2023 CHI Conference on Human Factors in Computing Systems}} (Hamburg, Germany) \emph{(\bibinfo{series}{CHI EA '23})}. \bibinfo{publisher}{Association for Computing Machinery}, \bibinfo{address}{New York, NY, USA}, Article \bibinfo{articleno}{63}, \bibinfo{numpages}{7}~pages.
\newblock
\showISBNx{9781450394222}
\urldef\tempurl%
\url{https://doi.org/10.1145/3544549.3585656}
\showDOI{\tempurl}


\bibitem[Seaborn et~al\mbox{.}(2023b)]%
        {234}
\bibfield{author}{\bibinfo{person}{Katie Seaborn}, \bibinfo{person}{Somang Nam}, \bibinfo{person}{Julia Keckeis}, {and} \bibinfo{person}{Tatsuya Itagaki}.} \bibinfo{year}{2023}\natexlab{b}.
\newblock \showarticletitle{Can Voice Assistants Sound Cute? Towards a Model of Kawaii Vocalics}. In \bibinfo{booktitle}{\emph{Extended Abstracts of the 2023 CHI Conference on Human Factors in Computing Systems}} (Hamburg, Germany) \emph{(\bibinfo{series}{CHI EA '23})}. \bibinfo{publisher}{Association for Computing Machinery}, \bibinfo{address}{New York, NY, USA}, Article \bibinfo{articleno}{63}, \bibinfo{numpages}{7}~pages.
\newblock
\showISBNx{9781450394222}
\urldef\tempurl%
\url{https://doi.org/10.1145/3544549.3585656}
\showDOI{\tempurl}


\bibitem[Seaborn et~al\mbox{.}(2023c)]%
        {1}
\bibfield{author}{\bibinfo{person}{Katie Seaborn}, \bibinfo{person}{Katja Rogers}, \bibinfo{person}{Somang Nam}, {and} \bibinfo{person}{Miu Kojima}.} \bibinfo{year}{2023}\natexlab{c}.
\newblock \showarticletitle{Kawaii Game Vocalics: A Preliminary Model}. In \bibinfo{booktitle}{\emph{Companion Proceedings of the Annual Symposium on Computer-Human Interaction in Play}} (Stratford, ON, Canada) \emph{(\bibinfo{series}{CHI PLAY Companion '23})}. \bibinfo{publisher}{Association for Computing Machinery}, \bibinfo{address}{New York, NY, USA}, \bibinfo{pages}{202–208}.
\newblock
\showISBNx{9798400700293}
\urldef\tempurl%
\url{https://doi.org/10.1145/3573382.3616099}
\showDOI{\tempurl}


\bibitem[Shinohara et~al\mbox{.}(2023)]%
        {30}
\bibfield{author}{\bibinfo{person}{Maino Shinohara}, \bibinfo{person}{Daisuke Sakamoto}, \bibinfo{person}{Tetsuo Ono}, {and} \bibinfo{person}{James~Everett Young}.} \bibinfo{year}{2023}\natexlab{}.
\newblock \showarticletitle{Understanding Privacy-friendly Design of Robot Eyes}. In \bibinfo{booktitle}{\emph{Proceedings of the 11th International Conference on Human-Agent Interaction}} (Gothenburg, Sweden) \emph{(\bibinfo{series}{HAI '23})}. \bibinfo{publisher}{Association for Computing Machinery}, \bibinfo{address}{New York, NY, USA}, \bibinfo{pages}{133–141}.
\newblock
\showISBNx{9798400708244}
\urldef\tempurl%
\url{https://doi.org/10.1145/3623809.3623829}
\showDOI{\tempurl}


\bibitem[Shinohara et~al\mbox{.}(2018)]%
        {94}
\bibfield{author}{\bibinfo{person}{Yumiko Shinohara}, \bibinfo{person}{Kazuhiro Mitsukuni}, \bibinfo{person}{Takayuki Yoneda}, \bibinfo{person}{Jun Ichikawa}, \bibinfo{person}{Yukiko Nishizaki}, {and} \bibinfo{person}{Natsuki Oka}.} \bibinfo{year}{2018}\natexlab{}.
\newblock \showarticletitle{A Humanoid Robot Can Use Mimicry to Increase Likability and Motivation for Helping}. In \bibinfo{booktitle}{\emph{Proceedings of the 6th International Conference on Human-Agent Interaction}} (Southampton, United Kingdom) \emph{(\bibinfo{series}{HAI '18})}. \bibinfo{publisher}{Association for Computing Machinery}, \bibinfo{address}{New York, NY, USA}, \bibinfo{pages}{122–128}.
\newblock
\showISBNx{9781450359535}
\urldef\tempurl%
\url{https://doi.org/10.1145/3284432.3284437}
\showDOI{\tempurl}


\bibitem[Shiomi and Hagita(2016)]%
        {44}
\bibfield{author}{\bibinfo{person}{Masahiro Shiomi} {and} \bibinfo{person}{Norihiro Hagita}.} \bibinfo{year}{2016}\natexlab{}.
\newblock \showarticletitle{Do Synchronized Multiple Robots Exert Peer Pressure?}. In \bibinfo{booktitle}{\emph{Proceedings of the Fourth International Conference on Human Agent Interaction}} (Biopolis, Singapore) \emph{(\bibinfo{series}{HAI '16})}. \bibinfo{publisher}{Association for Computing Machinery}, \bibinfo{address}{New York, NY, USA}, \bibinfo{pages}{27–33}.
\newblock
\showISBNx{9781450345088}
\urldef\tempurl%
\url{https://doi.org/10.1145/2974804.2974808}
\showDOI{\tempurl}


\bibitem[Sone(2016)]%
        {sone2016japanese}
\bibfield{author}{\bibinfo{person}{Yuji Sone}.} \bibinfo{year}{2016}\natexlab{}.
\newblock \bibinfo{booktitle}{\emph{Japanese robot culture}}.
\newblock \bibinfo{publisher}{Springer}, \bibinfo{address}{London, UK}.
\newblock


\bibitem[Sripian et~al\mbox{.}(2020)]%
        {173}
\bibfield{author}{\bibinfo{person}{Peeraya Sripian}, \bibinfo{person}{Kejkaew Thanasuan}, \bibinfo{person}{Keiko Miyatake}, {and} \bibinfo{person}{Michiko Ohkura}.} \bibinfo{year}{2020}\natexlab{}.
\newblock \showarticletitle{The Analysis of Kawaii Fashion in Thailand and Japan Using Colorfulness Metrics}. In \bibinfo{booktitle}{\emph{Advances in Affective and Pleasurable Design}}, \bibfield{editor}{\bibinfo{person}{Shuichi Fukuda}} (Ed.). \bibinfo{publisher}{Springer International Publishing}, \bibinfo{address}{Cham}, \bibinfo{pages}{224--231}.
\newblock
\showISBNx{978-3-030-20441-9}


\bibitem[Suegami et~al\mbox{.}(2017)]%
        {70}
\bibfield{author}{\bibinfo{person}{Takashi Suegami}, \bibinfo{person}{Hidenobu Sumioka}, \bibinfo{person}{Fuminao Obayashi}, \bibinfo{person}{Kyonosuke Ichii}, \bibinfo{person}{Yoshinori Harada}, \bibinfo{person}{Hiroshi Daimoto}, \bibinfo{person}{Aya Nakae}, {and} \bibinfo{person}{Hiroshi Ishiguro}.} \bibinfo{year}{2017}\natexlab{}.
\newblock \showarticletitle{Endocrinological Responses to a New Interactive HMI for a Straddle-type Vehicle: A Pilot Study}. In \bibinfo{booktitle}{\emph{Proceedings of the 5th International Conference on Human Agent Interaction}} (Bielefeld, Germany) \emph{(\bibinfo{series}{HAI '17})}. \bibinfo{publisher}{Association for Computing Machinery}, \bibinfo{address}{New York, NY, USA}, \bibinfo{pages}{463–467}.
\newblock
\showISBNx{9781450351133}
\urldef\tempurl%
\url{https://doi.org/10.1145/3125739.3132588}
\showDOI{\tempurl}


\bibitem[Suzuki and Umemuro(2012)]%
        {202}
\bibfield{author}{\bibinfo{person}{Daisuke Suzuki} {and} \bibinfo{person}{Hiroyuki Umemuro}.} \bibinfo{year}{2012}\natexlab{}.
\newblock \showarticletitle{Dimensions of people's attitudes toward robots}. In \bibinfo{booktitle}{\emph{Proceedings of the Seventh Annual ACM/IEEE International Conference on Human-Robot Interaction}} (Boston, Massachusetts, USA) \emph{(\bibinfo{series}{HRI '12})}. \bibinfo{publisher}{Association for Computing Machinery}, \bibinfo{address}{New York, NY, USA}, \bibinfo{pages}{249–250}.
\newblock
\showISBNx{9781450310635}
\urldef\tempurl%
\url{https://doi.org/10.1145/2157689.2157779}
\showDOI{\tempurl}


\bibitem[Tricco et~al\mbox{.}(2018)]%
        {Tricco2018}
\bibfield{author}{\bibinfo{person}{Andrea~C. Tricco}, \bibinfo{person}{Erin Lillie}, \bibinfo{person}{Wasifa Zarin}, \bibinfo{person}{Kelly~K. O’Brien}, \bibinfo{person}{Heather Colquhoun}, \bibinfo{person}{Danielle Levac}, \bibinfo{person}{David Moher}, \bibinfo{person}{Micah~D.J. Peters}, \bibinfo{person}{Tanya Horsley}, \bibinfo{person}{Laura Weeks}, \bibinfo{person}{Susanne Hempel}, \bibinfo{person}{Elie~A. Akl}, \bibinfo{person}{Christine Chang}, \bibinfo{person}{Jessie McGowan}, \bibinfo{person}{Lesley Stewart}, \bibinfo{person}{Lisa Hartling}, \bibinfo{person}{Adrian Aldcroft}, \bibinfo{person}{Michael~G. Wilson}, \bibinfo{person}{Chantelle Garritty}, \bibinfo{person}{Simon Lewin}, \bibinfo{person}{Christina~M. Godfrey}, \bibinfo{person}{Marilyn~T. Macdonald}, \bibinfo{person}{Etienne~V. Langlois}, \bibinfo{person}{Karla Soares-Weiser}, \bibinfo{person}{Jo Moriarty}, \bibinfo{person}{Tammy Clifford}, \bibinfo{person}{\"{O}zge Tun\c{c}alp}, {and} \bibinfo{person}{Sharon~E. Straus}.}
  \bibinfo{year}{2018}\natexlab{}.
\newblock \showarticletitle{PRISMA Extension for Scoping Reviews (PRISMA-ScR): Checklist and Explanation}.
\newblock \bibinfo{journal}{\emph{Annals of Internal Medicine}} \bibinfo{volume}{169}, \bibinfo{number}{7} (\bibinfo{date}{Oct.} \bibinfo{year}{2018}), \bibinfo{pages}{467–473}.
\newblock
\showISSN{1539-3704}
\urldef\tempurl%
\url{https://doi.org/10.7326/m18-0850}
\showDOI{\tempurl}


\bibitem[Tsuji et~al\mbox{.}(2019)]%
        {150}
\bibfield{author}{\bibinfo{person}{Amato Tsuji}, \bibinfo{person}{Keita Ushida}, \bibinfo{person}{Saneyasu Yamaguchi}, {and} \bibinfo{person}{Qiu Chen}.} \bibinfo{year}{2019}\natexlab{}.
\newblock \showarticletitle{Real-Time Collaborative Animation of 3D Models with Finger Play and Hand Shadow}. In \bibinfo{booktitle}{\emph{2019 IEEE Conference on Virtual Reality and 3D User Interfaces (VR)}}. \bibinfo{publisher}{IEEE}, \bibinfo{address}{New York, NY, USA}, \bibinfo{pages}{1195--1196}.
\newblock
\urldef\tempurl%
\url{https://doi.org/10.1109/VR.2019.8797847}
\showDOI{\tempurl}


\bibitem[Xu et~al\mbox{.}(2014)]%
        {219}
\bibfield{author}{\bibinfo{person}{Xin Xu}, \bibinfo{person}{Jianming Wu}, \bibinfo{person}{Kengo Fujita}, \bibinfo{person}{Tsuneo Kato}, {and} \bibinfo{person}{Fumiaki Sugaya}.} \bibinfo{year}{2014}\natexlab{}.
\newblock \showarticletitle{Hey Peratama: a breeding game with spoken dialogue interface}. In \bibinfo{booktitle}{\emph{Proceedings of the 13th International Conference on Mobile and Ubiquitous Multimedia}} (Melbourne, Victoria, Australia) \emph{(\bibinfo{series}{MUM '14})}. \bibinfo{publisher}{Association for Computing Machinery}, \bibinfo{address}{New York, NY, USA}, \bibinfo{pages}{266–267}.
\newblock
\showISBNx{9781450333047}
\urldef\tempurl%
\url{https://doi.org/10.1145/2677972.2678011}
\showDOI{\tempurl}


\bibitem[Yamamoto et~al\mbox{.}(2014)]%
        {64}
\bibfield{author}{\bibinfo{person}{Daisuke Yamamoto}, \bibinfo{person}{Keiichiro Oura}, \bibinfo{person}{Ryota Nishimura}, \bibinfo{person}{Takahiro Uchiya}, \bibinfo{person}{Akinobu Lee}, \bibinfo{person}{Ichi Takumi}, {and} \bibinfo{person}{Keiichi Tokuda}.} \bibinfo{year}{2014}\natexlab{}.
\newblock \showarticletitle{Voice interaction system with 3D-CG virtual agent for stand-alone smartphones}. In \bibinfo{booktitle}{\emph{Proceedings of the Second International Conference on Human-Agent Interaction}} (Tsukuba, Japan) \emph{(\bibinfo{series}{HAI '14})}. \bibinfo{publisher}{Association for Computing Machinery}, \bibinfo{address}{New York, NY, USA}, \bibinfo{pages}{323–330}.
\newblock
\showISBNx{9781450330350}
\urldef\tempurl%
\url{https://doi.org/10.1145/2658861.2658874}
\showDOI{\tempurl}


\bibitem[Yamazaki et~al\mbox{.}(2019)]%
        {66}
\bibfield{author}{\bibinfo{person}{Ryuji Yamazaki}, \bibinfo{person}{Hiroko Kase}, \bibinfo{person}{Shuichi Nishio}, {and} \bibinfo{person}{Hiroshi Ishiguro}.} \bibinfo{year}{2019}\natexlab{}.
\newblock \showarticletitle{A Conversational Robotic Approach to Dementia Symptoms: Measuring Its Effect on Older Adults}. In \bibinfo{booktitle}{\emph{Proceedings of the 7th International Conference on Human-Agent Interaction}} (Kyoto, Japan) \emph{(\bibinfo{series}{HAI '19})}. \bibinfo{publisher}{Association for Computing Machinery}, \bibinfo{address}{New York, NY, USA}, \bibinfo{pages}{110–117}.
\newblock
\showISBNx{9781450369220}
\urldef\tempurl%
\url{https://doi.org/10.1145/3349537.3351888}
\showDOI{\tempurl}


\bibitem[Yeh et~al\mbox{.}(2017)]%
        {99}
\bibfield{author}{\bibinfo{person}{Alexander Yeh}, \bibinfo{person}{Photchara Ratsamee}, \bibinfo{person}{Kiyoshi Kiyokawa}, \bibinfo{person}{Yuki Uranishi}, \bibinfo{person}{Tomohiro Mashita}, \bibinfo{person}{Haruo Takemura}, \bibinfo{person}{Morten Fjeld}, {and} \bibinfo{person}{Mohammad Obaid}.} \bibinfo{year}{2017}\natexlab{}.
\newblock \showarticletitle{Exploring Proxemics for Human-Drone Interaction}. In \bibinfo{booktitle}{\emph{Proceedings of the 5th International Conference on Human Agent Interaction}} (Bielefeld, Germany) \emph{(\bibinfo{series}{HAI '17})}. \bibinfo{publisher}{Association for Computing Machinery}, \bibinfo{address}{New York, NY, USA}, \bibinfo{pages}{81–88}.
\newblock
\showISBNx{9781450351133}
\urldef\tempurl%
\url{https://doi.org/10.1145/3125739.3125773}
\showDOI{\tempurl}


\bibitem[Yomota(2006)]%
        {yomota2006kawaii}
\bibfield{author}{\bibinfo{person}{Inuhiko Yomota}.} \bibinfo{year}{2006}\natexlab{}.
\newblock \bibinfo{booktitle}{\emph{"Kawaii" ron}}.
\newblock \bibinfo{publisher}{Chikuma Shobo}, \bibinfo{address}{Tokyo, Japan}.
\newblock


\bibitem[Zhang et~al\mbox{.}(2021)]%
        {zhang2021exploring}
\bibfield{author}{\bibinfo{person}{Brian~J Zhang}, \bibinfo{person}{Knut Peterson}, \bibinfo{person}{Christopher~A Sanchez}, {and} \bibinfo{person}{Naomi~T Fitter}.} \bibinfo{year}{2021}\natexlab{}.
\newblock \showarticletitle{Exploring Consequential Robot Sound: Should We Make Robots Quiet and Kawaii-et?}. In \bibinfo{booktitle}{\emph{2021 IEEE/RSJ International Conference on Intelligent Robots and Systems (IROS)}}. \bibinfo{publisher}{IEEE}, \bibinfo{address}{New York, NY, USA}, \bibinfo{pages}{3056--3062}.
\newblock


\end{thebibliography}

\end{document}